\documentclass[preprint,12pt,authoryear]{elsarticle}
\usepackage{graphicx}
\usepackage{multirow}
\usepackage{amsmath,amssymb,amsfonts}
\usepackage{amsthm}
\usepackage{mathrsfs}
\usepackage{orcidlink}
\usepackage{rotating}
\usepackage[title]{appendix}
\usepackage{xcolor}
\usepackage{textcomp}
\usepackage{manyfoot}
\usepackage{tabularx}
\usepackage{booktabs}
\usepackage{ltablex}
\usepackage{algorithm}
\usepackage{makecell}
\usepackage{geometry}
\usepackage{array}
\usepackage{algorithmicx}
\usepackage{algpseudocode}
\usepackage{listings}
\usepackage{placeins}
\usepackage{lineno,hyperref}
\hypersetup{colorlinks,
citecolor=blue,colorlinks=true,linkcolor=blue, urlcolor=blue, linktocpage,
hypertexnames=false
}
\usepackage[T1]{fontenc}

\usepackage{float}

\usepackage{pgfplots}
\pgfplotsset{compat=1.18}
\usepackage{pgfplotstable}
\usepackage{tikz}
\usetikzlibrary{patterns}
\usetikzlibrary{mindmap,shadows}
\usetikzlibrary{arrows.meta,positioning,fit,calc,shapes.geometric}
\usepgfplotslibrary{groupplots}

\usepackage{bm}
\newcommand{\vect}[1]{\bm{#1}}

\journal{Computer Science Review}
\begin{document}
\begin{frontmatter}
\title{Towards Efficient and Secure Cloud-Assisted Autonomous Systems: A Review of Architectures, Algorithms, Security, and Deployment Challenges}
\author[mymainaddress2,mymainaddress1]{Yasir Ali\orcidlink{0000-0003-3998-9218}}
\author[mymainaddress]{Tayyab Manzoor\orcidlink{0000-0003-3932-0506}}
\author[mymainaddress1]{Huan Yang\orcidlink{0000-0001-8539-7104}}
\author[mymainaddress1]{Asif Ali\orcidlink{0000-0001-6895-8897}}
\author[mymainaddress3,mymainaddress1]{Yuanqing Xia\orcidlink{0000-0002-5977-4911}\texorpdfstring{\corref{cor1}}{}}
\ead{xia\_yuanqing@bit.edu.cn}
\cortext[cor1]{Corresponding author}

\address[mymainaddress2]{Zhengzhou Research Institute, Beijing Institute of Technology, Zhengzhou 450000, Henan Province, China}
\address[mymainaddress1]{School of Automation, Beijing Institute of Technology, Beijing 100081, China}
\address[mymainaddress]{School of Automation and Electrical Engineering, Zhongyuan University of Technology, Zhengzhou 450007, Henan Province, China}
\address[mymainaddress3]{Zhongyuan University of Technology, Zhengzhou 450007, Henan Province, China}
\begin{abstract}
Cloud Control Systems (CCSs) are emerging as distributed computing infrastructures for autonomous and semi-autonomous cyber-physical systems, where sensing, estimation, optimization, learning, security monitoring, and recovery may be executed across local devices, edge/fog nodes, and cloud services. This review synthesizes CCS research from 2012 to 2026 for computer science audiences, emphasizing how cloud-edge-fog computing, distributed algorithms, data-driven intelligence, cybersecurity, privacy-preserving computation, and runtime assurance reshape the design of closed-loop intelligent systems. Rather than treating CCSs as conventional Networked Control Systems (NCSs) with remote computation, the review frames them as cyber-physical service architectures whose correctness depends on timing-aware task placement, dependable middleware, secure data and model flows, resilient decision pipelines, local fallback, and verifiable deployment evidence. We review cloud-assisted Model Predictive Control (MPC), workflow and scheduling methods, encrypted and privacy-preserving control, learning-enabled CCSs, Denial-of-Service (DoS) and False Data Injection (FDI) resilience, audit and governance mechanisms, hybrid cloud-edge-fog architectures, and testbed or field evidence in robotics, industrial automation, transportation, energy, and human-facing autonomy. The synthesis identifies open problems in benchmark design, real-time guarantees under variable cloud execution, trustworthy AI update pipelines, security-by-design, privacy-performance trade-offs, fallback certification, and safety-case documentation. The review concludes with a research agenda for CCSs as secure, scalable, and verifiable computing infrastructures for trustworthy autonomous cyber-physical systems.
\end{abstract}
\begin{keyword}
Cloud control systems\sep cloud-edge-fog computing\sep cybersecurity and privacy\sep distributed algorithms\sep trustworthy intelligent systems.
\end{keyword}
\end{frontmatter}
\section{Introduction}\label{sec1}
Modern autonomous and industrial cyber-physical systems (CPSs) are becoming distributed, data-intensive, and service-connected \citep{DOOSTMOHAMMADIAN2025100983, zhang2019networked}. Mobile robots and Unmanned Aerial Vehicles (UAVs) operate as fleets, factories expose production assets through Industrial Internet of Things (IIoT) platforms, energy systems coordinate large numbers of flexible resources, and transportation networks increasingly combine local vehicle autonomy with infrastructure-level prediction \citep{kehoe2015cloudrobotics, wang2021cloud, li2022secure, smartgrids1, smarttransfor2}. These systems require control decisions over multiple time scales: millisecond stabilization, second-level safety supervision, minute-level optimization, and long-horizon model learning \citep{zhang2019networked, xia2024cloud}. The result is a mismatch between the growing computational and data-management demands of autonomy and the limited resources, maintenance capacity, and security posture of many local controllers \citep{DOOSTMOHAMMADIAN2025100983, zhang2019networked}.

NCSs enabled remote feedback and distributed sensing by connecting sensors, controllers, and actuators via communication networks \citep{zhang2019networked}. CCSs inherit this foundation but change the architecture and trust model more deeply. In a CCS, control-relevant intelligence may be delegated to elastic cloud services, regional fog nodes, edge gateways, or local embedded controllers \citep{13-xia2012networked, xia2015cloud, xia2024cloud}. The cloud may solve long-horizon MPC problems, train perception or prediction models, maintain digital twins, coordinate fleets, audit safety evidence, or distribute controller updates \citep{dai2023cloud, kehoe2015cloudrobotics, minerva2020digitaltwin, jin2025cloud}. Thus, a CCS is not simply an NCS plus remote computing. It is a closed-loop service infrastructure in which sensing, estimation, optimization, learning, privacy protection, security monitoring, and fallback recovery can be distributed across organizational and computational boundaries.

\begin{figure}[!htbp]
\centering
\includegraphics[width=\textwidth]{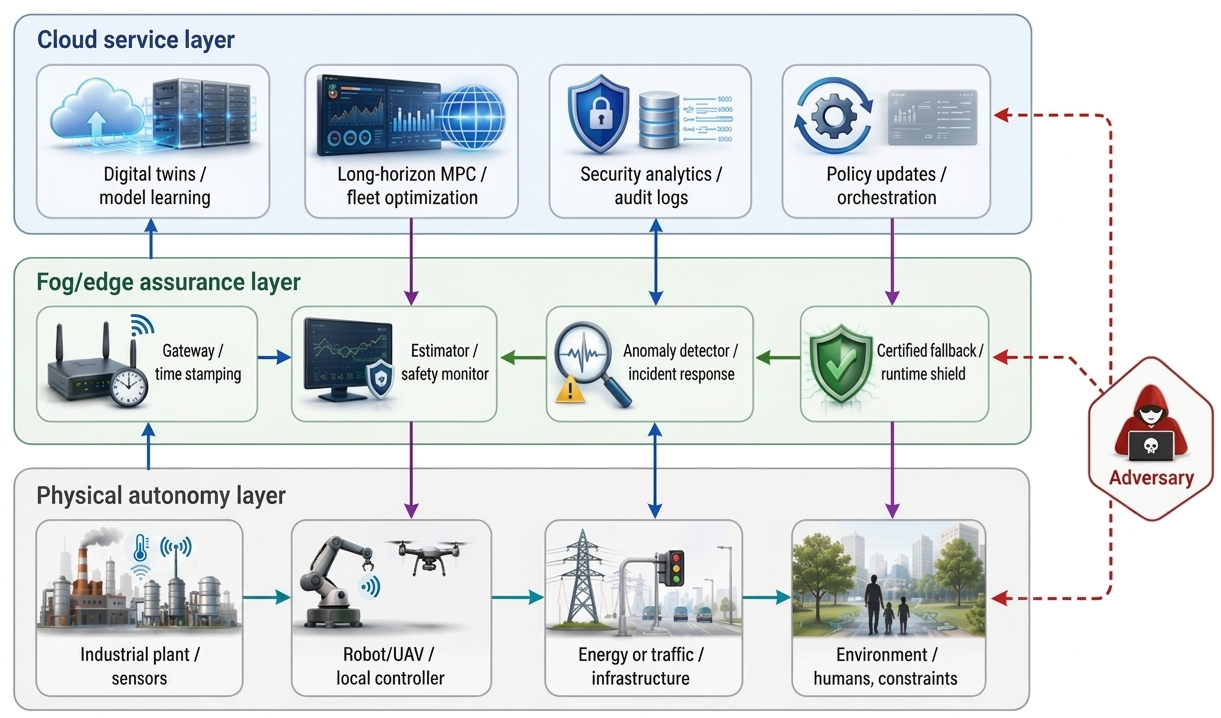}
\caption{Layered CCS architecture for trustworthy cyber-physical autonomy. The analytical point is task placement: cloud services provide model learning, long-horizon optimization, security analytics, and orchestration; fog and edge nodes provide timing-aware estimation, anomaly detection, and fallback; and local plants, robots, and infrastructure retain authority over fast safety-critical interactions.}
\label{fig:layered_ccs_architecture}
\end{figure}

Fig.~\ref{fig:layered_ccs_architecture} illustrates the layered CCS architecture that motivates this review. The figure emphasizes that practical CCSs are not single cloud controllers, but distributed cyber-physical service infrastructures in which cloud services, fog and edge nodes, local safety logic, and physical robotic or industrial assets interact under timing, communication, security, and resilience constraints. This layered view also clarifies why CCS design requires joint treatment of algorithms, distributed computing architecture, security and privacy, and deployment evidence.

This shift is timely because robotics and autonomy are moving from isolated prototypes toward persistent computational services. Cloud and edge infrastructure can provide scalable optimization, shared mapping, fleet-level learning, anomaly detection, digital-twin maintenance, and security-aware operational monitoring \citep{kehoe2015cloudrobotics, zhan2019future, minerva2020digitaltwin, jin2025cloud}. At the same time, these benefits come with systems-level risks: latency, jitter, packet loss, cloud outage, cloud-provider trust assumptions, privacy leakage, adversarial command or sensor manipulation, controller-version drift, and unclear responsibility when a remote service affects a physical plant. For safety-critical autonomy, the decisive question is not whether the cloud can compute a better policy in principle, but whether the entire cloud-edge-fog feedback service can operate within timing, safety, privacy, security, and recovery constraints.

In computer science, CCSs are important because they turn the control of physical systems into a distributed computing problem, with real-time, safety, privacy, security, and accountability constraints. Task placement, workflow scheduling, middleware quality-of-service, encrypted computation, model-update governance, runtime monitoring, and benchmark design become part of the correctness argument. A robot or industrial controller cannot be judged only by nominal tracking error if its policies, maps, diagnostics, or fallback decisions are updated through cloud services. Conversely, cloud robotics and industrial IoT cannot be evaluated only as computing architectures if cloud-hosted decisions affect closed-loop stability and safety. Fast local loops, runtime assurance, hybrid cloud-edge-fog allocation, privacy-preserving computation, resilient estimation, and evidence maturity are therefore inseparable CCS design issues \citep{schierman2020runtime, hsu2024safetyfilter}.

Prior reviews address important adjacent literatures, including secure NCSs, control under delay and packet loss, cloud robotics, CPS security, edge/fog computing for IoT, encrypted control, privacy-preserving optimization, and learning-enabled autonomous systems \citep{kehoe2015cloudrobotics, heemels2012eventtriggered, cardenas2008securecontrol, brunke2022safelearning}. These areas provide essential foundations, but they usually emphasize one part of the problem: secure NCSs emphasize attacks on networked feedback; delay-tolerant control emphasizes timing uncertainty; cloud robotics emphasizes shared computation and data services; CPS security emphasizes threats and defenses; edge/fog computing emphasizes resource placement; encrypted control emphasizes confidential computation; and learning-enabled autonomy emphasizes data-driven adaptation. The present review adds a closed-loop service perspective: it asks what happens when estimation, optimization, learning, security monitoring, audit, and recovery functions are delegated across local, edge, fog, and cloud layers. This is not merely another survey of cyberattacks, cloud computing, or networked control; it is a synthesis of the algorithmic, distributed-systems, security, privacy, control, and deployment consequences of that delegation.

\begin{table}[!htbp]
\centering
\caption{How this review differs from adjacent literatures}
\label{tab:adjacent_literatures}
\begingroup
\footnotesize
\setlength{\tabcolsep}{4pt}
\renewcommand{\arraystretch}{1.04}
\resizebox{0.97\textwidth}{!}{%
\begin{tabular}{p{4.2cm}p{5.8cm}p{8.4cm}}
\toprule
\textbf{Adjacent literature} & \textbf{Main emphasis} & \textbf{What this review adds} \\
\midrule
\textbf{Secure NCSs} & Attacks and defenses for feedback over communication networks & Treats the cloud, edge, fog, middleware, credentials, model repositories, and update pipelines as parts of the control service \\
\midrule
\textbf{Control under delay and packet loss} & Stability and performance under network timing uncertainty & Connects delay to cloud execution, orchestration, fallback authority, privacy overhead, and deployment evidence \\
\midrule
\textbf{Cloud robotics} & Shared computation, mapping, data, and learning services for robots & Explains which robotic autonomy functions can be cloud-hosted and which require local or edge safety authority \\
\midrule
\textbf{CPS security} & Threat models, intrusion detection, and resilient operation of cyber-physical assets & Links confidentiality, integrity, availability, and accountability to closed-loop stability, safety, and recovery \\
\midrule
\textbf{Edge/fog computing for IoT} & Resource placement, latency reduction, and distributed service architecture & Adds feedback-specific requirements: sampling deadlines, actuator authority, fallback switching, and safety-case evidence \\
\midrule
\textbf{Encrypted and privacy-preserving control} & Confidential computation and protected data sharing & Evaluates privacy mechanisms by their computation, communication, and closed-loop performance consequences \\
\midrule
\textbf{Learning-enabled autonomy} & Data-driven perception, prediction, adaptation, and policy learning & Frames cloud-trained or cloud-updated models as safety-relevant software artifacts requiring validation, rollback, and audit \\
\bottomrule
\end{tabular}}
\endgroup
\end{table}

Table~\ref{tab:adjacent_literatures} contrasts the present review with adjacent \emph{research areas}. Because \emph{Computer Science Review} readers will also ask how this article differs from existing \emph{survey} articles, Table~\ref{tab:prior_surveys} adds an explicit survey-to-survey comparison. Prior reviews typically advance a single viewpoint---networked-control security, cloud robotics, cloud-control architecture, safe learning, or edge/fog offloading---whereas the present review integrates these viewpoints through a single closed-loop service lens and grades the literature using an explicit evidence-maturity ladder.

\begin{table}[!htbp]
\centering
\caption{Positioning against representative prior surveys and reviews}
\label{tab:prior_surveys}
\begingroup
\footnotesize
\setlength{\tabcolsep}{4pt}
\renewcommand{\arraystretch}{1.05}
\resizebox{\textwidth}{!}{%
\begin{tabular}{p{2.8cm}p{3.0cm}p{3.0cm}p{3.2cm}p{4.0cm}}
\toprule
\textbf{Prior survey/review} & \textbf{Primary lens} & \textbf{Security/privacy depth} & \textbf{Deployment-evidence treatment} & \textbf{What this review adds} \\
\midrule
\textbf{Networked control systems} \citep{zhang2019networked} & Feedback and estimation over communication networks & Network attacks (DoS, FDI, replay) on feedback & Mainly analytical and simulation results & Cloud-edge-fog task placement, privacy-as-computation cost, and an evidence-maturity ladder \\
\midrule
\textbf{Cloud robotics} \citep{kehoe2015cloudrobotics} & Shared computation, mapping, data, and learning for robots & Limited; not centered on feedback security & Prototype and cloud-service demonstrations & A closed-loop timing contract, security-in-the-loop, and fallback certification \\
\midrule
\textbf{Cloud control overview} \citep{xia2022brief, xia2024cloud} & Cloud-based control concepts and architecture & Mentions security; limited privacy-computation depth & Conceptual plus selected testbeds & Evidence grading and a CS-facing scheduling, benchmark, and governance agenda \\
\midrule
\textbf{Safe learning for control} \citep{brunke2022safelearning} & Data-driven and safe learning guarantees & Robustness and safety, not cyber-security & Simulation and some hardware & Cloud delegation, encrypted/private computation, and model-update governance \\
\midrule
\textbf{Edge/fog computing} \citep{zhan2019future} & Resource placement and control offloading & Architectural security only & Mostly conceptual & Feedback deadline contract, certified fallback, and safety-case evidence \\
\midrule
\textbf{This review} & CCS as a closed-loop cyber-physical \emph{service} (2012--2026) & Confidentiality, integrity, availability, and accountability plus privacy as closed-loop costs & Explicit ladder from analytical to certified service & Unifies algorithms, task placement, security/privacy, and deployment evidence in one framework \\
\bottomrule
\end{tabular}}
\endgroup
\end{table}

The central thesis is that CCSs can become deployable infrastructure for trustworthy autonomous CPSs only if cloud intelligence is combined with timing-aware cloud-edge-fog allocation, certified local fallback, privacy-preserving computation, security-by-design, runtime assurance, and evidence maturity. The paper organizes the field around four dimensions: \emph{algorithmic and control intelligence}, \emph{cloud-edge-fog computing architecture}, \emph{security, privacy, and trust}, and \emph{evidence and deployment maturity}. The main take-home message is selective rather than encyclopedic: cloud assistance is most compelling for long-horizon, high-dimensional, multi-agent, data-intensive, or governance-heavy functions; fast safety-critical loops usually require local or edge authority; and deployment claims should be calibrated to the available evidence.

\begin{figure}[!htbp]
\centering
\includegraphics[width=\textwidth]{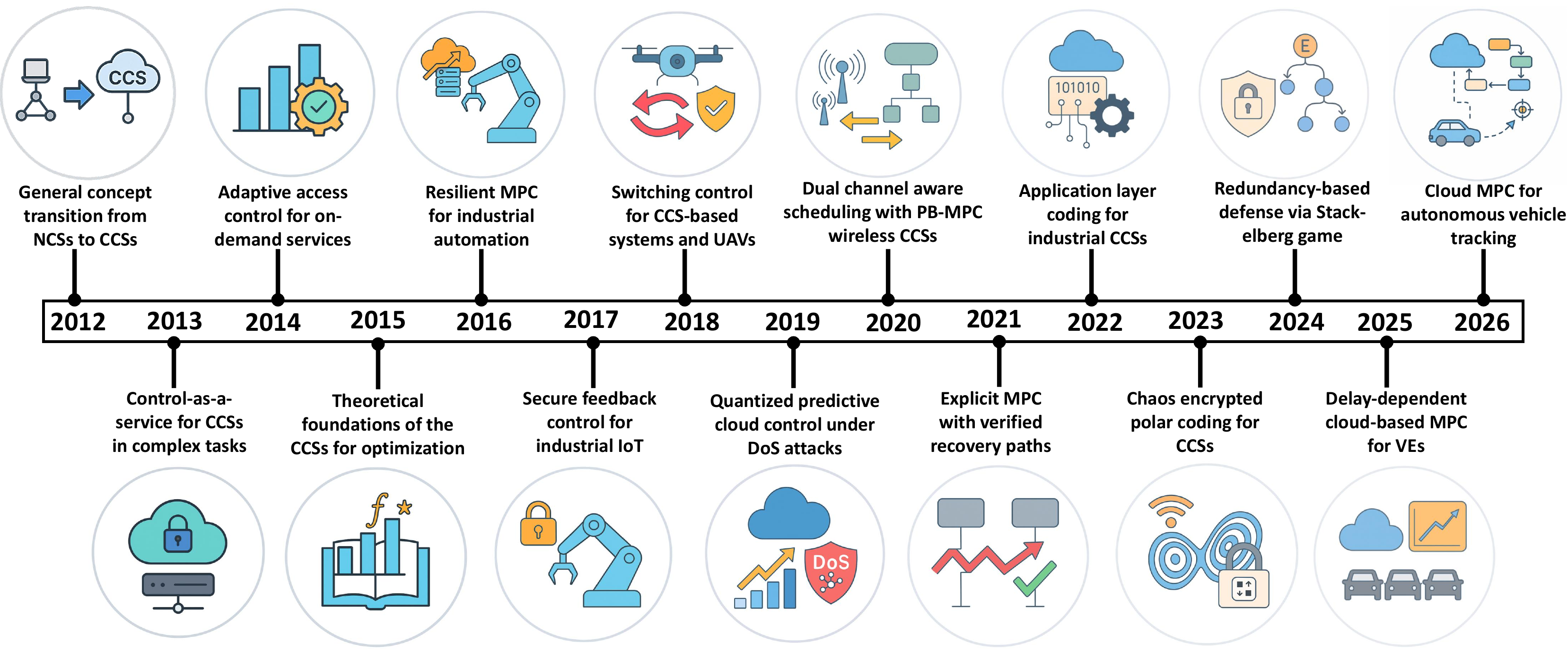}
\caption{Chronological evolution of CCS research from 2012 to 2026. The sequence illustrates a shift from cloud-enabled computation toward secure, resilient, privacy-aware, and deployment-oriented control-service design.}\label{Fig1new}
\end{figure}

Table~\ref{tab1} summarizes the distinction between NCSs and CCSs, while Fig.~\ref{Fig1new} traces the progression from early conceptual work to recent secure, resilient, and application-specific designs. The evolution of CCSs from 2012 to 2026 shows a movement from cloud-enabled computation toward trustworthy control service design. Early efforts focused on the conceptual transition from NCSs to CCSs \citep{13-xia2012networked}, adaptive access control for on-demand services \citep{younis2014access}, resilient and switching MPC for industrial automation and UAVs \citep{LEI2016324, skarin9683307}, and packet-based MPC for wireless CCSs \citep{li2020integrated}. Later work added application-layer security coding \citep{Yuan9987498}, DoS-resilient architectures \citep{yang2019predictive}, explicit MPC with verified recovery paths \citep{skarin9683307}, string-stable delay-dependent cloud predictive control for vehicle platoons \citep{zhorfei10879591}, and workflow-based real-time cloud MPC for autonomous-vehicle trajectory tracking \citep{gao2026cloudenabled}. The field now needs synthesis around timing, assurance, privacy, and deployment evidence rather than a longer catalogue of isolated techniques.
\begin{table}[!htbp]
  \centering
  \caption{NCSs and CCSs viewed through control-service consequences}
  \label{tab1}
  \resizebox{0.9\textwidth}{!}{ \begin{tabular}{p{3.5cm}p{4cm}p{8cm}}
    \toprule
    \textbf{Property} & \textbf{NCSs} & \textbf{CCSs} \\
    \midrule
    \textbf{Location of intelligence} & Feedback law is commonly assigned to a local or remote networked controller & Estimation, optimization, learning, supervision, audit, and fallback may be split across local, edge, fog, and cloud layers \\
    \textbf{Timing problem} & Delay and packet loss are network-induced constraints on feedback & Timing includes sensing, uplink, queueing, cloud solve time, downlink, actuation, orchestration, and service availability \\
    \textbf{Control objective} & Maintain stability/performance over a communication network & Maintain stability, safety, privacy, security, recoverability, and accountability for a control service \\
    \textbf{Scalability} & Limited by local controllers and communication topology & Strong for fleet coordination, long-horizon optimization, data-driven modeling, and lifecycle monitoring when timing permits \\
    \textbf{Trust model} & Trust is centered on the plant owner and communication network & Trust extends to cloud providers, software supply chains, model repositories, API access, and cross-tenant data protection \\
    \textbf{Failure mode} & Network degradation or node failure affects the feedback loop & Cloud outage, service misconfiguration, delayed optimization, unsafe model updates, and compromised credentials can affect physical behavior \\
    \textbf{Deployment evidence} & Stability and robustness analyses are central & Stability remains necessary but must be supported by delay traces, cloud execution data, fallback validation, security evidence, and operational governance \\
    \bottomrule
  \end{tabular}}
\end{table}
The contribution of this review is to connect computational architecture, distributed algorithms, security and privacy mechanisms, control methodology, autonomous-CPS use cases, and deployment maturity within one interpretive framework. The aim is not to claim that CCSs replace local control, nor to suggest that every control computation should move to the cloud. Instead, the review asks a more useful decision question for computer science and cyber-physical autonomy: which functions can be cloud-hosted, which must remain local or edge-resident, and what evidence is required before a cloud-assisted design can be trusted in physical operation?

\noindent The specific contributions of this review are as follows.
\begin{itemize}
\item A closed-loop \emph{service} framework that treats sensing, estimation, optimization, learning, security monitoring, audit, and recovery as functions distributed across local, edge, fog, and cloud layers, rather than as an NCS with remote computation (Section~\ref{sec:framework}).
\item A function-level placement analysis that maps control and autonomy functions to suitable computational layers together with their timing, security, privacy, and assurance requirements (Table~\ref{tab:function_placement}).
\item A unified treatment of confidentiality, integrity, availability, and accountability as closed-loop properties, with encrypted MPC, DoS resilience, FDI mitigation, and auditability evaluated by their control consequences rather than as isolated security tools (Section~\ref{sec:security_trust}).
\item An evidence-maturity ladder (Table~\ref{tab:evidence_maturity}) that calibrates deployment claims from analytical results to certified services, applied consistently throughout the review.
\item A computer-science-facing research agenda covering real-time cloud control, privacy-preserving feedback, trustworthy AI pipelines, hybrid orchestration, and a reproducible CCS benchmark-reporting template (Table~\ref{tab:benchmark_requirements}).
\end{itemize}

Algorithmic and control intelligence captures the evolution from model-based MPC and LQG toward data-driven, learning-enabled, and fleet-level autonomy methods. The cloud-edge-fog computing architecture captures the shift from centralized cloud control to hybrid service placement. Security, privacy, and trust capture privacy-preserving computation, encrypted control, auditability, DoS resilience, and FDI mitigation. Evidence and deployment maturity distinguish analytical proofs, simulations, hardware-in-the-loop experiments, cloud/edge testbeds, field trials, and certified or audited operational services. Table~\ref{tab:review_framework} summarizes this review logic.

The growing need for hybrid control systems that integrate CCSs with edge and fog computing has become especially evident. Each paradigm offers distinct advantages, as shown in Table~\ref{tab:paradigm_comparison}. CCSs provide elastic scalability and centralized optimization, but they may not be suitable for time-sensitive tasks due to latency, unpredictable availability, and centralized attack surfaces. Edge computing provides shorter response paths, making it appropriate for fast local dynamics such as robotics, while fog computing provides an intermediate layer that supports regional coordination, context awareness, and partial resilience. No single paradigm dominates all control scenarios, so future CCSs are likely to be hybrid, with time-critical safety loops at the edge, coordination and filtering at fog nodes, and long-horizon optimization, learning, and fleet-level intelligence in the cloud \citep{zhan2019future, boiko2024edge, jin2025cloud}.

Specifically, the review addresses three focal questions:

\begin{itemize}
\item What control, estimation, optimization, learning, and fleet-level autonomy functions are suitable for cloud assistance, and under what timing assumptions?
\item How do confidentiality, integrity, availability, accountability, and privacy mechanisms alter closed-loop performance and safety?
\item What deployment evidence is needed to move from promising CCS algorithms to trustworthy services for industrial, robotic, transportation, energy, and human-facing autonomous systems?
\end{itemize}

\begin{figure}[!ht]
\centering
\includegraphics[width=0.75\textwidth]{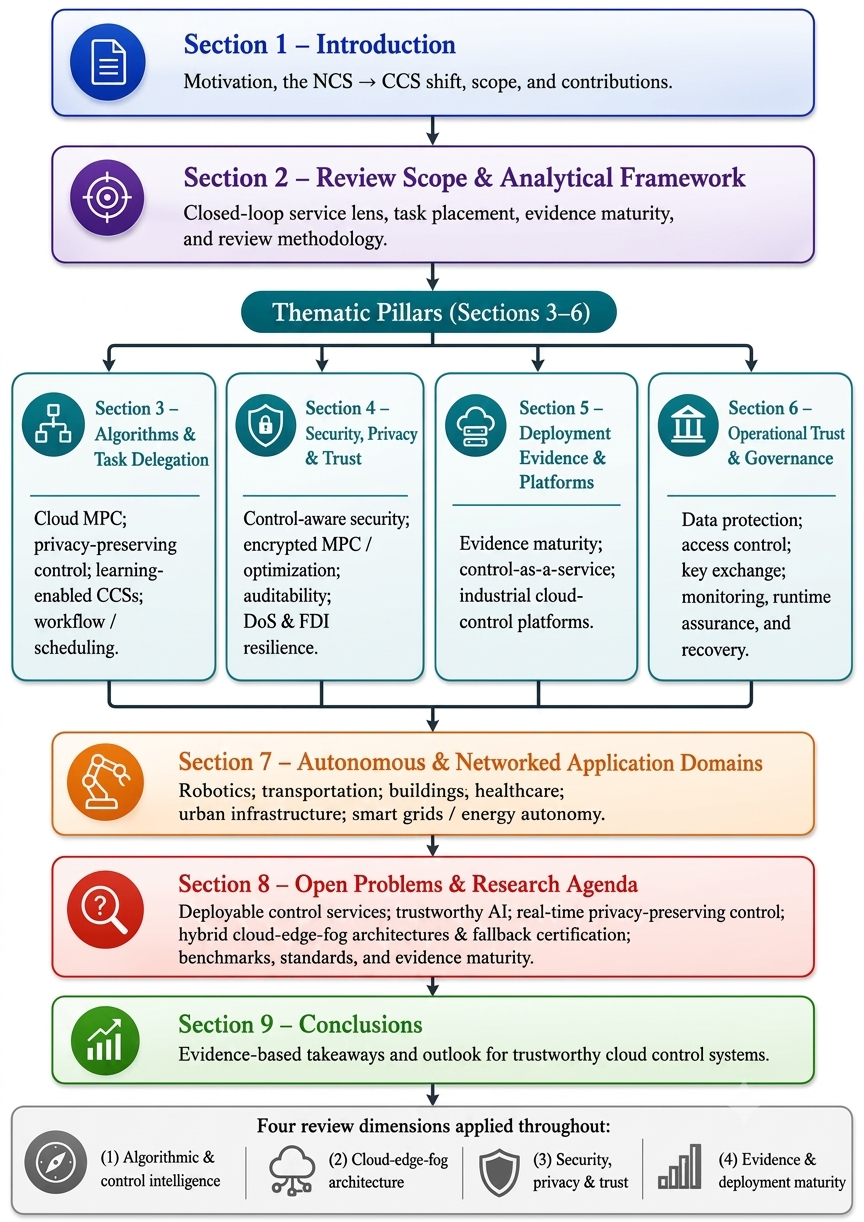}
\caption{Organization and logical flow of the review.}
\label{fig:paper_organization}
\end{figure}

The remainder of the paper is organized as follows. Section~\ref{sec:framework} defines the review scope, notation, and closed-loop service framework. Section~\ref{sec:control_intelligence} reviews computational algorithms and task delegation across the cloud-edge continuum, including cloud-assisted MPC, privacy-preserving control, secure optimization, and learning-enabled CCSs. Section~\ref{sec:security_trust} examines security, privacy, and trust as feedback-loop properties, with emphasis on encrypted MPC, auditability, DoS resilience, and FDI mitigation. Section~\ref{sec:deployment_maturity} discusses deployment evidence and cloud-control platforms, while Section~\ref{sec:operational_trust} addresses operational trust, governance, data protection, recovery, patch management, and human factors. Section~\ref{sec:application_domains} synthesizes autonomous and networked application domains, including robotics, transportation, healthcare, smart cities, and energy autonomy. Section~\ref{sec:research_agenda} then develops the open research agenda and grand challenges, including real-time cloud control, trustworthy AI, hybrid architectures, fallback certification, benchmarks, and standardization. The paper concludes with key findings and final conclusions. Fig.~\ref{fig:paper_organization} summarizes this organization and the logical flow of the review.

\begin{table}[!htbp]
\centering
\caption{Computer-science-facing framework used in this review}
\label{tab:review_framework}
\resizebox{0.97\textwidth}{!}{
\begin{tabular}{p{4cm}p{5.5cm}p{5.8cm}p{5.2cm}}
\toprule
\textbf{Dimension} & \textbf{Computer science question} & \textbf{Representative topics} & \textbf{Evidence required} \\
\midrule
\textbf{Algorithmic and control intelligence} & Which estimation, optimization, learning, and coordination algorithms can be safely provided as remote or distributed services? & MPC, LQG, distributed optimization, data-driven control, learning-enabled autonomy & Timing traces, feasibility or stability evidence, fallback behavior, and update validation \\
\midrule
\textbf{Cloud-edge-fog computing architecture} & How should tasks, data, models, and authority be placed across local devices, edge/fog nodes, and cloud services? & Cloud-only control, edge-cloud systems, fog coordination, middleware, workflow scheduling & Placement rationale, resource limits, QoS data, outage behavior, and orchestration overhead \\
\midrule
\textbf{Security, privacy, and trust} & How do confidentiality, integrity, availability, accountability, and privacy mechanisms change closed-loop service correctness? & Encrypted MPC, homomorphic encryption, differential privacy, secure logs, DoS and FDI mitigation & Threat model, overhead, recovery behavior, auditability, and effect on stability or safety \\
\midrule
\textbf{Evidence and deployment maturity} & What evidence distinguishes an algorithmic demonstration from a dependable cyber-physical computing service? & Simulations, testbeds, cloud platforms, field trials, standards, safety fallback & Reproducible benchmarks, lifecycle evidence, operator workflow, certification or audit pathway \\
\bottomrule
\end{tabular}}
\end{table}
\begin{table}[!htbp]
\centering
\caption{Cloud, edge, and fog computing paradigms for CCS task placement}
\label{tab:paradigm_comparison}
 \resizebox{\textwidth}{!}{
\begin{tabular}{p{5cm}p{5cm}p{5cm}p{5cm}}
\toprule
\textbf{Aspect} & \textbf{Cloud Control Systems} & \textbf{Edge Computing} & \textbf{Fog Computing} \\
\midrule
\textbf{Primary Deployment} &
Remote data processing centers (centralized cloud infrastructure) &
Local devices or gateways (e.g., PLCs, IoT gateways) &
Intermediate nodes between edge and cloud (e.g., routers, micro-data centers) \\
\midrule
\textbf{Typical control placement} &
Long-horizon optimization, model learning, digital twins, fleet coordination, audit logging, and supervisory decisions when deadlines permit &
Fast feedback, safety monitors, local perception, fallback controllers, and time-critical inference near the plant &
Regional coordination, event filtering, cooperative estimation, local caching, and intermediate assurance for groups of assets \\
\midrule
\textbf{Timing implication} &
Variable round-trip delay and server-side jitter; suitable when the decision horizon is longer than communication and solve time &
Shortest response path; needed for stabilization, collision avoidance, emergency stops, and fast robot or vehicle dynamics &
Intermediate response path; useful when coordination needs more context than a single edge node but cannot wait for remote cloud decisions \\
\midrule
\textbf{Data Processing} &
Raw sensor data is transmitted to the cloud for centralized processing. High bandwidth required. &
Local processing at the data source. Only aggregated/events are sent upstream. Bandwidth-efficient. &
Distributed processing across fog nodes. Pre-processed data is relayed to the cloud. Enables near-real-time decision-making with regional context awareness. \\
\midrule
\textbf{Scalability} &
High computational and storage elasticity, with dependence on network and cloud-provider availability &
Constrained by device resources but resilient to cloud disconnection &
Moderate and distributed; scales through regional nodes but requires orchestration and topology management \\
\midrule
\textbf{Security Implications} &
\textbullet{} Vulnerable to Man-in-the-Middle (MitM) attacks \newline
\textbullet{} Requires strong encryption and cloud access governance \newline
\textbullet{} Centralized attack surface \newline
\textbullet{} Complex compliance (e.g., GDPR, HIPAA) &
\textbullet{} Physically accessible attack vectors \newline
\textbullet{} Lightweight local protection and hardware roots of trust \newline
\textbullet{} Local trust zones \newline
\textbullet{} Susceptible to tampering if physical access is gained &
\textbullet{} Distributed attack surface \newline
\textbullet{} Requires end-to-end encryption \newline
\textbullet{} Blockchain integration feasible for audit trails \newline
\textbullet{} Trusted Execution Environments (TEEs) can enhance privacy \\
\midrule
\textbf{Failure resilience} &
Unsafe if the plant depends on cloud-only feedback; viable when paired with predictive buffers, redundant services, and verified local fallback &
Best location for fail-safe and degraded operation under network or cloud outage &
Supports graceful degradation through local caching, rerouting, and regional fallback when cloud connectivity is intermittent \\
\midrule
\textbf{Use Case Examples} &
\textbullet{} Predictive control services \newline
\textbullet{} Large-scale smart-grid optimization \newline
\textbullet{} Historical data analytics \newline
\textbullet{} Long-horizon production planning \& management &
\textbullet{} Anti-lock braking systems \newline
\textbullet{} Industrial servo control \newline
\textbullet{} Real-time anomaly detection \newline
\textbullet{} Autonomous drones and mobile robots &
\textbullet{} Traffic-light coordination \newline
\textbullet{} Smart-building automation \newline
\textbullet{} Distributed sensor networks \newline
\textbullet{} Urban mobility management \\
\bottomrule
\end{tabular}}
\end{table}

\section{Review Scope and Analytical Framework}\label{sec:framework}
\subsection{Review Scope and Selection Logic}
This article is a critical narrative review rather than a bibliometric survey. Its purpose is to synthesize the computing-systems, algorithmic, security/privacy, and autonomous-CPS implications of CCSs, not merely to count publications. The review covers foundational CCS and networked-control studies from approximately 2012 onward, together with recent work through 2026 on cloud-based MPC, privacy-preserving optimization, encrypted control, DoS and FDI resilience, auditability mechanisms, cloud-edge-fog architectures, runtime operation, and industrial or autonomous-system deployments \citep{zhang2019networked, 13-xia2012networked, xia2022brief, xia2024cloud, dai2023cloud, gao2026cloudenabled, chen2026clouddistributed}. The 2012 starting point is appropriate because the modern CCS concept began to differentiate from classical NCSs with the emergence of cloud computing as an explicit control service resource \citep{13-xia2012networked}.

This review prioritizes studies that connect cloud, edge, or fog computation to feedback control, estimation, optimization, security, privacy, or runtime operation in cyber-physical or autonomous systems \citep{zhan2019future, jin2025cloud, kehoe2015cloudrobotics}. Generic cloud computing, generic IoT cybersecurity, and big-data analytics papers are treated as contextual unless they have an explicit closed-loop consequence. Central topics include cloud-assisted MPC and optimization, encrypted and privacy-preserving control, cloud-edge-fog control architecture, cyber-resilient feedback, autonomous-system deployment, and evidence maturity \citep{mayne2000constrainedmpc, alexandru2020cloud, naseri2022encrypted, feng2023secure, yang2019predictive, yin2022cloud}. Contextual topics include general cloud infrastructure, generic access control, and broad IoT analytics when they clarify but do not define the control problem.

The evidence evaluation is intentionally cautious. Analytical work is assessed for assumptions and guarantees; simulations are assessed for plant models, timing assumptions, and threat models; testbeds are assessed for end-to-end timing and integration evidence; and field trials are assessed for operational reliability, fallback behavior, and governance \citep{mayne2000constrainedmpc, quevedo2011packetized, heemels2012eventtriggered, schierman2020runtime}. The review is narrative because the literature spans heterogeneous control problems, cloud platforms, robotic systems, cryptographic mechanisms, and application domains that do not yet share common benchmarks. A purely bibliometric count would obscure the central issue: whether a claimed CCS contribution changes closed-loop operation in a trustworthy and deployable way.

The scope has limitations. The review is not a complete bibliometric census, and some emerging 2026 contributions may still be at the preprint stage or only partially validated. Cross-study quantitative comparison is limited because authors report plant dynamics, sampling periods, delay distributions, cloud execution times, safety metrics, privacy assumptions, and attack models inconsistently. Deployment claims are also difficult to compare because field trials, cloud/edge testbeds, and safety cases are reported with different levels of detail. These limitations motivate the benchmark-and-evidence agenda developed later in the paper.
\subsection{Review Methodology}
To keep the selection reproducible while preserving the narrative character of the review, we followed a structured but deliberately non-exhaustive protocol. Literature was retrieved from IEEE Xplore, Scopus, the ACM Digital Library, ScienceDirect, and Web of Science, supplemented by backward and forward citation tracing (snowballing) from foundational cloud-control and CPS-security papers and by recent preprints for 2025--2026 contributions. Search strings combined a control/computing axis (\emph{cloud control}, \emph{networked control}, \emph{cloud-based MPC}, \emph{edge/fog control}, \emph{workflow scheduling}, \emph{digital twin}) with a trust axis (\emph{encrypted control}, \emph{privacy-preserving}, \emph{differential privacy}, \emph{denial of service}, \emph{false data injection}, \emph{runtime assurance}, \emph{safe learning}).

Studies were included when they satisfied three criteria: (i) an explicit closed-loop or feedback consequence of cloud, edge, or fog computation; (ii) treatment of at least one review dimension---algorithmic intelligence, architecture and task placement, security and privacy, or deployment evidence; and (iii) publication in a peer-reviewed venue or a citable preprint with sufficient methodological detail. Studies were excluded when they addressed generic cloud computing, IoT, or big-data analytics without a control consequence, presented position statements without technical or empirical content, or duplicated superseded versions. The search window spans 2012--2026, with 2012 chosen because the modern CCS concept began to separate from classical NCSs around the emergence of cloud computing as an explicit control-service resource \citep{13-xia2012networked}.

After de-duplication and screening against these criteria, the synthesis retained the studies cited throughout Sections~\ref{sec:control_intelligence}--\ref{sec:application_domains}, prioritizing work that changes closed-loop behavior in a trustworthy and deployable way over work that adds incremental algorithmic variation. Consistent with a critical narrative review, the retained studies were grouped thematically and graded using the evidence-maturity levels defined in Table~\ref{tab:evidence_maturity} rather than pooled quantitatively, because heterogeneous plants, sampling periods, threat models, and reporting conventions preclude meta-analytic comparison. Fig.~\ref{fig:methodology} summarizes this protocol.

\begin{figure}[!htbp]
\centering
\resizebox{0.9\textwidth}{!}{%
\begin{tikzpicture}[
  font=\sffamily,
  >=Latex,
  line cap=round, line join=round,
  stage/.style={rectangle, rounded corners=3pt, line width=0.8pt, align=center,
    text width=8.6cm, inner sep=7pt, drop shadow={opacity=0.2}},
  incl/.style={rectangle, rounded corners=3pt, draw=green!45!black, fill=green!9,
    line width=0.8pt, align=left, text width=4.7cm, inner sep=6pt, drop shadow={opacity=0.2}},
  excl/.style={rectangle, rounded corners=3pt, draw=red!60!black, fill=red!7,
    line width=0.8pt, align=left, text width=4.7cm, inner sep=6pt, drop shadow={opacity=0.2}},
  flow/.style={->, line width=1.0pt, draw=black!55}
]
\node[stage, draw=blue!55!black, fill=blue!8] (id) at (0,0)
  {\textbf{Identification.} Bibliographic databases --- IEEE~Xplore, Scopus, ACM~Digital Library, ScienceDirect, and Web of Science --- supplemented by backward/forward citation tracing (snowballing) and 2025--2026 preprints};
\node[stage, draw=violet!60!black, fill=violet!8] (search) at (0,-3.6)
  {\textbf{Search strategy.} Control/computing axis (\emph{cloud control}, \emph{cloud-based MPC}, \emph{edge/fog control}, \emph{workflow scheduling}, \emph{digital twin}) \textbf{AND} trust axis (\emph{encrypted control}, \emph{differential privacy}, \emph{denial of service}, \emph{false data injection}, \emph{runtime assurance}, \emph{safe learning})};
\node[stage, draw=orange!75!black, fill=orange!10] (screen) at (0,-6.7)
  {\textbf{Screening \& eligibility.} Title/abstract and full-text assessment against the inclusion and exclusion criteria};
\node[stage, draw=teal!60!black, fill=teal!9] (syn) at (0,-9.1)
  {\textbf{Synthesis.} De-duplication $\rightarrow$ retained corpus (2012--2026) $\rightarrow$ thematic grouping $\rightarrow$ evidence-maturity grading (Table~\ref{tab:evidence_maturity})};
\node[incl] (inc) at (-8.1,-6.7)
  {\textbf{Include if}\\ (i) explicit closed-loop / feedback consequence;\\ (ii) $\ge 1$ review dimension (algorithms, architecture, security/privacy, evidence);\\ (iii) peer-reviewed venue or citable preprint};
\node[excl] (exc) at (8.1,-6.7)
  {\textbf{Exclude if}\\ generic cloud / IoT / big-data without a control consequence;\\ position papers without technical content;\\ duplicate or superseded versions};
\draw[flow] (id) -- (search);
\draw[flow] (search) -- (screen);
\draw[flow] (screen) -- (syn);
\draw[->, line width=0.9pt, draw=green!45!black] (inc.east) -- (screen.west);
\draw[->, line width=0.9pt, draw=red!60!black, dashed] (screen.east) -- (exc.west);
\end{tikzpicture}%
}
\caption{Review methodology. Studies are identified from major bibliographic databases and supplemented by snowballing and recent preprints, retrieved through a control/computing axis combined with a trust axis, screened against explicit inclusion and exclusion criteria, and synthesized thematically with evidence-maturity grading (Table~\ref{tab:evidence_maturity}) rather than pooled quantitatively.}
\label{fig:methodology}
\end{figure}
\subsection*{Notation}
Throughout the paper, bold lowercase symbols denote vectors, bold uppercase
symbols denote matrices when used, calligraphic uppercase symbols denote sets,
operators, function classes, or mechanisms, and non-bold Greek or Roman symbols
denote scalars unless stated otherwise. The subscript $k$ denotes discrete time,
$0\!:\!k$ denotes a sequence from time $0$ to $k$, and superscripts such as
$a$, $y$, and $u$ denote attacked, measurement-channel, and input-channel
quantities when used in the security model.
\subsection{Closed-Loop Service Framework}
The analytical unit used throughout the review is the closed-loop service, not the cloud server alone. Building on standard state-space and networked-control representations, a CCS can be represented as a plant--network--controller interconnection in which sensing, estimation, optimization, and supervision may be distributed across local, edge, fog, and cloud layers \citep{zhang2019networked, 13-xia2012networked, xia2024cloud, zhan2019future}:
\begin{equation}
\begin{aligned}
\vect{x}_{k+1} &= f(\vect{x}_k,\vect{u}_k,\vect{w}_k),
\qquad \vect{y}_k = h(\vect{x}_k,\vect{v}_k),\\
\hat{\vect{x}}_k &= \mathcal{E}_{\ell_k}\!\left(
\vect{y}_{0:k},\tau_{0:k},\gamma_{0:k},\mathcal{I}_{0:k}
\right),\\
\vect{u}_k &= \mathcal{K}_{\ell_k}\!\left(
\hat{\vect{x}}_k,\vect{r}_k,\vect{\theta}_k,\mathcal{S}_k
\right),
\end{aligned}
\label{eq:ccs_closed_loop}
\end{equation}
where $\vect{x}_k$ is the plant state vector, $\hat{\vect{x}}_k$ its estimate, $\vect{u}_k$ the control-input vector, $\vect{y}_k$ the measured-output vector, $\vect{r}_k$ the reference or set-point vector, $\vect{w}_k$ and $\vect{v}_k$ process and measurement disturbance vectors, $\vect{\theta}_k$ the controller or model parameters (e.g., cloud-updated gains or learned weights), $\mathcal{E}_{\ell_k}$ and $\mathcal{K}_{\ell_k}$ the estimator and controller maps evaluated on layer $\ell_k$, $\ell_k\in\{\mathrm{local},\mathrm{edge},\mathrm{fog},\mathrm{cloud}\}$ the computational layer selected at time $k$, $\tau_k$ the communication and computation delay, $\gamma_k$ the packet-delivery indicator, $\mathcal{I}_{0:k}$ the information available to the remote service, and $\mathcal{S}_k$ the security or privacy mechanism active in the loop.

From this perspective, the design problem is not simply to maximize cloud computation. A mature CCS must allocate control functions across layers while satisfying performance, timing, privacy, and resilience constraints. The following compact formulation is inspired by constrained MPC, networked predictive control, cloud-based MPC, privacy-preserving cloud control, and resilient control under network attacks \citep{mayne2000constrainedmpc, quevedo2011packetized}:
\begin{equation}
\begin{aligned}
\min_{\mathcal{K},\mathcal{E},\ell_{0:N}} \quad &
\mathbb{E}\!\left[
\sum_{k=0}^{N}\ell_x(\vect{x}_k,\vect{r}_k)+\ell_u(\vect{u}_k)
\right] + C_{\mathrm{net}} + C_{\mathrm{sec}} \\
\mathrm{s.t.}\quad &
\vect{x}_{k+1}=f(\vect{x}_k,\vect{u}_k,\vect{w}_k),\quad
\vect{x}_k\in\mathcal{X},\quad \vect{u}_k\in\mathcal{U},\\
& \Pr(\tau_k \leq T_s)\geq 1-\epsilon,\quad
\mathcal{L}_{\mathrm{privacy}}\leq \bar{\mathcal{L}},\quad
\mathcal{R}_{\mathrm{attack}}\geq \bar{\mathcal{R}}.
\end{aligned}
\label{eq:ccs_design_problem}
\end{equation}
where $T_s$ is the sampling period or decision deadline, $\mathcal{L}_{\mathrm{privacy}}$ denotes an information-leakage or privacy-loss measure, and $\mathcal{R}_{\mathrm{attack}}$ denotes resilience under attacks or outages. The formulation in~\eqref{eq:ccs_design_problem} is not intended as a single universal formulation for all CCSs. It is a compact way to expose the main trade-offs that organize the literature: deeper cloud optimization can improve nominal performance, but only if timing, confidentiality, integrity, availability, and fallback behavior are handled as control constraints rather than afterthoughts \citep{alexandru2020secure, feng2023secure, schierman2020runtime, hsu2024safetyfilter}. Table~\ref{tab:tradeoffs} distils the trade-offs that Eq.~\eqref{eq:ccs_design_problem} encodes into concrete placement rules; the remaining sections substantiate each row with specific mechanisms and evidence.
\begin{table}[!htbp]
\centering
\caption{Cross-layer co-design trade-offs in CCSs and the resulting placement rules}
\label{tab:tradeoffs}
\begingroup
\footnotesize
\setlength{\tabcolsep}{4pt}
\renewcommand{\arraystretch}{1.1}
\resizebox{\textwidth}{!}{%
\begin{tabular}{>{\raggedright\arraybackslash}p{2.8cm}>{\raggedright\arraybackslash}p{3.0cm}>{\raggedright\arraybackslash}p{2.9cm}>{\raggedright\arraybackslash}p{2.7cm}>{\raggedright\arraybackslash}p{4.4cm}}
\toprule
\textbf{Trade-off axis} & \textbf{Gained by ``more''} & \textbf{Paid in} & \textbf{Binding constraint} & \textbf{Placement rule of thumb} \\
\midrule
\textbf{Cloud optimization depth} & Longer horizons, nonlinear and fleet-level optimization, better nominal performance & Round-trip delay, jitter, and outage exposure & Deadline $\Pr(\tau_k\le T_s)\ge 1-\epsilon$ & Deepen the cloud role only while the deadline holds with margin; otherwise demote the cloud to references, constraints, or policy updates consumed by a local fallback \\
\midrule
\textbf{Privacy / encryption strength} & Confidentiality of states, models, and constraints & Per-step computation, ciphertext bandwidth, and quantization error & Sampling-period feasibility & Apply strong cryptography on slow or medium-rate loops; use lightweight or edge-side protection on fast loops, and report timing alongside privacy strength \\
\midrule
\textbf{Integrity and resilience redundancy} & Tolerance to FDI and DoS, graceful degradation & Compute, memory, and verification or design cost & Edge resource budget & Size redundancy to the worst tolerated outage or attack window; certify the fallback \emph{switch}, not only the nominal loop \\
\midrule
\textbf{Centralized cloud availability} & Elastic scale and shared data or learning & Single-region outage and cloud-provider trust dependence & Service availability $\gamma_k$ & Pair any cloud-critical function with verified local or edge fallback; never let cloud-only feedback own a safety-critical decision \\
\midrule
\textbf{Learning / model-update freshness} & Adaptivity, better prediction and diagnosis & Validation, uncertainty quantification, and rollback governance & Distribution-shift and safety budget & Gate every update behind validation, safety shielding, and rollback; treat learned components as safety-relevant artifacts \\
\midrule
\textbf{Accountability / audit logging} & Integrity, non-repudiation, and multi-party evidence & Consensus latency and storage overhead & Feedback-path latency budget & Keep ledgers off the fast loop; use them for post-event accountability, evidence sharing, and governance \\
\bottomrule
\end{tabular}}
\endgroup
\end{table}

To avoid overinterpreting isolated demonstrations, the review also distinguishes evidence maturity. A method supported only by simulation can be scientifically valuable, but it cannot carry the same deployment claim as a field-tested or safety-certified cloud-control service. Table~\ref{tab:evidence_maturity} defines the evidence language used in later sections.
\begin{table}[!htbp]
\centering
\caption{Evidence maturity levels for CCS research}
\label{tab:evidence_maturity}
\begingroup
\footnotesize
\setlength{\tabcolsep}{3pt}
\renewcommand{\arraystretch}{0.88}
\resizebox{0.9\textwidth}{!}{%
\begin{tabular}{>{\raggedright\arraybackslash}p{2.5cm}>{\raggedright\arraybackslash}p{4.1cm}>{\raggedright\arraybackslash}p{4.1cm}>{\raggedright\arraybackslash}p{4.2cm}}
\toprule
\textbf{Level} & \textbf{Typical evidence} & \textbf{What it supports} & \textbf{What remains unproven} \\
\midrule
\textbf{Analytical} & Proofs of stability, convergence, privacy, or security & Principle feasibility and sufficient conditions & Network realism, overhead, and robustness beyond assumptions \\
\textbf{Simulation} & Numerical plant, attack scenario, or cloud-emulation study & Comparative behavior and parameter sensitivity & Hardware timing, cloud variability, operator workflow, and reliability \\
\makecell[l]{\textbf{Prototype/}\\\textbf{testbed}} & HIL, robotic platform, PLC testbed, cloud container, or edge gateway & End-to-end timing, communication load, integration, and failure observation & Long-duration reliability, multi-site scalability, and certification evidence \\
\textbf{Field trial} & Real plant, vehicle, robot, industrial process, or infrastructure deployment & Feasibility under uncontrolled operating conditions & Generalizability, safety-case completeness, and lifecycle governance \\
\textbf{Certified service} & Audited deployment with documented fallback, security, update, and recovery processes & Industrial adoption pathway and accountable operation & Transferability to faster dynamics, new threats, or different regulatory contexts \\
\bottomrule
\end{tabular}}
\endgroup
\end{table}
Table~\ref{tab:function_placement} translates the closed-loop service framework into a function-level placement guide. It separates functions that normally require local or edge authority, such as inner-loop stabilization, fast estimation, and fallback recovery, from functions that are better suited to fog or cloud assistance, such as fleet coordination, digital-twin updates, perception-model updates, anomaly detection, and audit logging. The table also emphasizes that placement is not only a latency decision: each function carries different security, privacy, and assurance requirements.

\begin{table}[!htbp]
\centering
\caption{Control and autonomy functions mapped to CCS layers, timing, risks, and assurance evidence}
\label{tab:function_placement}
\resizebox{0.9\textwidth}{!}{
\begin{tabular}{p{3.4cm}p{3.2cm}p{3.1cm}p{4.1cm}p{4.2cm}p{3.7cm}}
\toprule
\textbf{Control or autonomy function} & \textbf{Suitable layer} & \textbf{Timing sensitivity} & \textbf{Security/privacy concern} & \textbf{Required assurance evidence} & \textbf{Example applications} \\
\midrule
\textbf{Inner-loop stabilization} & Local controller or edge device & Very high & Command tampering, sensor spoofing, actuator compromise & Stability proof, worst-case latency bound, verified emergency stop and fallback & Robot joints, UAV attitude, vehicle stability \\
\midrule
\textbf{State estimation and filtering} & Local/edge for fast loops; fog/cloud for slower fusion & High to moderate & FDI attacks, leakage of sensitive states, replay attacks & Secure observer validation, residual thresholds, delay/packet-loss traces & Mobile robots, industrial processes, smart grids \\
\midrule
\textbf{MPC and trajectory optimization} & Edge/fog/cloud depending on deadline & Moderate to high & Model/constraint confidentiality, delayed or stale commands & Deadline inequality, recursive feasibility, fallback MPC, cloud execution traces & Autonomous vehicles, energy systems, production scheduling \\
\midrule
\textbf{Fleet coordination} & Fog and cloud & Moderate & Cross-agent data leakage, compromised coordination commands & Multi-agent safety constraints, communication-loss tests, recovery behavior & UAV swarms, warehouse robots, connected vehicles \\
\midrule
\textbf{Perception model updates} & Cloud training; edge deployment & Low to moderate & Training-data privacy, model poisoning, unsafe updates & Validation set reporting, distribution-shift tests, staged rollout and rollback & Cloud robotics, visual inspection, smart infrastructure \\
\midrule
\textbf{Anomaly detection and diagnostics} & Edge/fog/cloud & Moderate & False alarms, stealthy attacks, telemetry leakage & Attack model, detection delay, control impact, operator workflow & Industrial plants, healthcare monitoring, grid protection \\
\midrule
\textbf{Digital twin update} & Cloud with edge/fog synchronization & Low to moderate & Proprietary model leakage, stale twin state & Model-version control, synchronization error, audit trail & Manufacturing, EV charging, autonomous infrastructure \\
\midrule
\textbf{Fallback and recovery} & Local/edge with cloud audit & Very high during faults & Unsafe handoff, compromised recovery policy & Runtime assurance case, switchback logic, outage and attack drills & Robotics, transportation, process safety \\
\midrule
\textbf{Audit and safety-case logging} & Fog/cloud governance layer & Low for logging, high for incident response & Integrity, non-repudiation, access governance & Tamper-resistant logs, controller version record, responsibility boundary & Multi-operator infrastructure, regulated autonomy \\
\bottomrule
\end{tabular}}
\end{table}

\noindent The implications for the CCS framework are clear: task placement is a distributed-systems design decision with direct control consequences. Cloud resources are most valuable when they provide computation, memory, cross-system data integration, or governance capabilities that local devices cannot support. Conversely, they are least appropriate when unbounded latency or service dependence could compromise the plant's ability to maintain safe operation. The remainder of this review is organized around this framework. The next section examines how computational algorithms and control intelligence are delegated to cloud resources. Subsequent sections review confidentiality, integrity, availability, and accountability mechanisms as protections for control loops; assess industrial platforms and operational trust; synthesize future research directions and application domains; and conclude with key findings and open problems.
\section{Computational Algorithms and Task Delegation Across the Cloud-Edge Continuum}
\label{sec:control_intelligence}
Cloud-based control schemes can be broadly categorized by the function delegated to the cloud and by the assumptions made about communication, computation, and trust \citep{13-xia2012networked, xia2015cloud, xia2024cloud}. From a computational-systems perspective, the key question is not simply whether a cloud server can solve a larger optimization problem than an embedded controller. The deeper question is which parts of the sensing-estimation-optimization-actuation pipeline can be safely outsourced while preserving closed-loop stability, constraint satisfaction, privacy, and real-time responsiveness. This section reviews the main algorithms used in CCSs and highlights the trade-offs among algorithmic performance, system architecture, and deployment feasibility.
\subsection{Cloud-Based Model Predictive Control}
Model Predictive Control (MPC) is one of the most natural candidates for cloud execution because it repeatedly solves constrained optimization problems and can benefit from elastic computation \citep{dai2023cloud, 1-ali2018secure}. Cloud-based MPC is attractive for large plants, multi-agent systems, transportation networks, and energy systems where local embedded devices cannot easily solve high-dimensional or nonlinear programs within the sampling period \citep{mayne2000constrainedmpc, dai2023cloud, gao2026cloudenabled, chen2026clouddistributed}. Its value, however, depends on a delicate balance: the cloud can expand the feasible optimization horizon and model complexity, but round-trip delay, packet loss, and server-side variability can erode the very performance gains that motivate cloud deployment.

A useful way to state the cloud-MPC feasibility condition is through the decision deadline
\begin{equation}
\tau^{\mathrm{sense}}_k+\tau^{\mathrm{up}}_k+\tau^{\mathrm{solve}}_k+\tau^{\mathrm{down}}_k+\tau^{\mathrm{apply}}_k \leq T_s,
\label{eq:cloud_mpc_deadline}
\end{equation}
where the terms denote sensing, uplink communication, cloud solution time, downlink communication, and actuator-side application delay. If the inequality cannot be guaranteed with high probability, the cloud should not be responsible for the inner feedback loop. Instead, the remote optimizer should provide preview trajectories, terminal sets, policy updates, or supervisory decisions that a local or edge controller can safely refine. This distinction is central for robotics and autonomous systems, where millisecond-level stabilization often remains local while fleet-level coordination and long-horizon planning can be cloud-assisted.

\begin{figure}[!htbp]
\centering
\includegraphics[width=\textwidth]{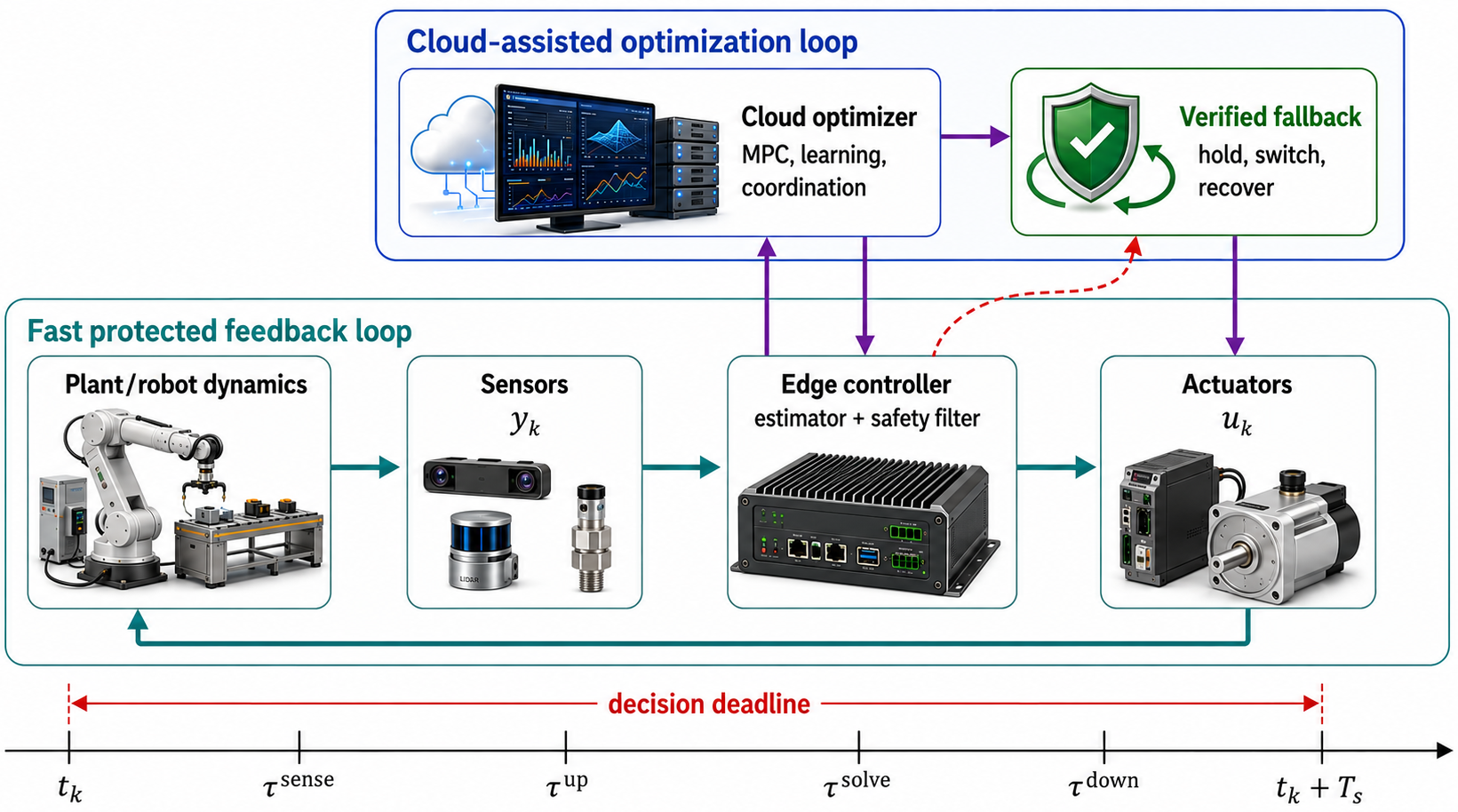}
\caption{Timing-aware cloud-MPC loop. Cloud optimization can participate directly in feedback only when sensing, uplink, cloud solution, downlink, and actuator-side application fit within the decision deadline. If the deadline is violated or uncertain, cloud outputs should be treated as references, constraints, or supervisory updates while the edge controller switches to a verified fallback policy.}
\label{fig:cloud_mpc_timing}
\end{figure}

Fig.~\ref{fig:cloud_mpc_timing} illustrates this timing-aware cloud-MPC contract and the fallback behavior triggered when the deadline is violated, uncertain, or only intermittently satisfied.

The literature illustrates several strategies for managing this balance. \citet{liu2020coordinated} studied coordinated control in networked multi-agent systems using distributed cloud computing, combining multi-step state prediction with coordination optimization to compensate for communication delays. \citet{dai2023cloud} proposed a cloud-oriented computational MPC scheme using a parallel multi-block Alternating Direction Method of Multipliers (ADMM) algorithm for nonlinear MPC with input nonlinearity, showing how decomposition and parallelism can make cloud execution more practical. \citet{gao2023predictive} applied a hierarchical cloud-based predictive cruise control architecture to urban buses at signalized intersections, where the cloud optimizes speed trajectories to reduce energy consumption and waiting time. More recent transportation-oriented CCS studies make the cloud-service issue explicit: \citet{gao2026cloudenabled} formulated workflow-based real-time MPC for autonomous vehicle trajectory tracking, while \citet{chen2026clouddistributed} proposed a cloud-based distributed data-enabled predictive-control framework for time-varying mixed traffic flow with communication-resilience strategies. Together, these examples show that cloud-based MPC is strongest when prediction horizons, fleet-level coordination, or nonlinear constraints exceed local computing capacity and weakest when hard real-time loops cannot tolerate uncertain communication delays.

The same deadline test yields a concrete placement rule for fast autonomous systems: attitude stabilization, collision avoidance, emergency stopping, and torque/servo loops should generally remain local or edge-resident, whereas cloud assistance is better suited to long-horizon planning, fleet-level coordination, energy-aware routing, or nonlinear optimization, whose benefit justifies the communication path. When the deadline holds only probabilistically, the cloud is best used to compute reference trajectories, terminal sets, tightened constraints, policy libraries, or digital twin updates for a local or edge controller, rather than to close the inner loop directly.

Several MPC variants are especially relevant to CCSs. Packetized predictive control sends a sequence of future inputs so the actuator can continue safely through short communication losses \citep{quevedo2011packetized, li2020integrated}. Event-triggered and adaptive triggering schemes reduce bandwidth and cloud calls, but they also create timing side channels and must be analyzed for stability under irregular updates \citep{heemels2012eventtriggered, yin2022cloud, yang2019predictive}. Explicit MPC and verified recovery controllers can provide local fallback when the cloud optimizer is late or unavailable \citep{skarin9683307}. Cloud-edge MPC and distributed cloud-fog designs attempt to split optimization across layers, using local feasibility and cloud-side long-horizon improvement \citep{guo2024cloud, li2024cooperative}. Data-enabled and sample-based predictive-control variants further broaden the design space when accurate models are unavailable, especially for safe motion planning and traffic systems \citep{huang2026sample, chen2026clouddistributed}. These variants inherit classical recursive-feasibility and stability concerns from constrained MPC \citep{mayne2000constrainedmpc}; in CCSs, those concerns must be rechecked under delayed, stale, missing, or rejected cloud updates. Anytime MPC is conceptually attractive because a cloud solver may return progressively improving solutions, but CCS studies should report what happens when the solver is interrupted, the returned solution is stale, or the edge controller rejects an unsafe update.

Implications for the CCS framework: cloud MPC is not a binary choice between local and remote control. It is a layered contract among optimization depth, communication timing, fallback authority, and safety evidence. The strongest CCS designs are those that make this contract explicit.
\begin{figure}[!t]
\centering
\includegraphics[width=\linewidth]{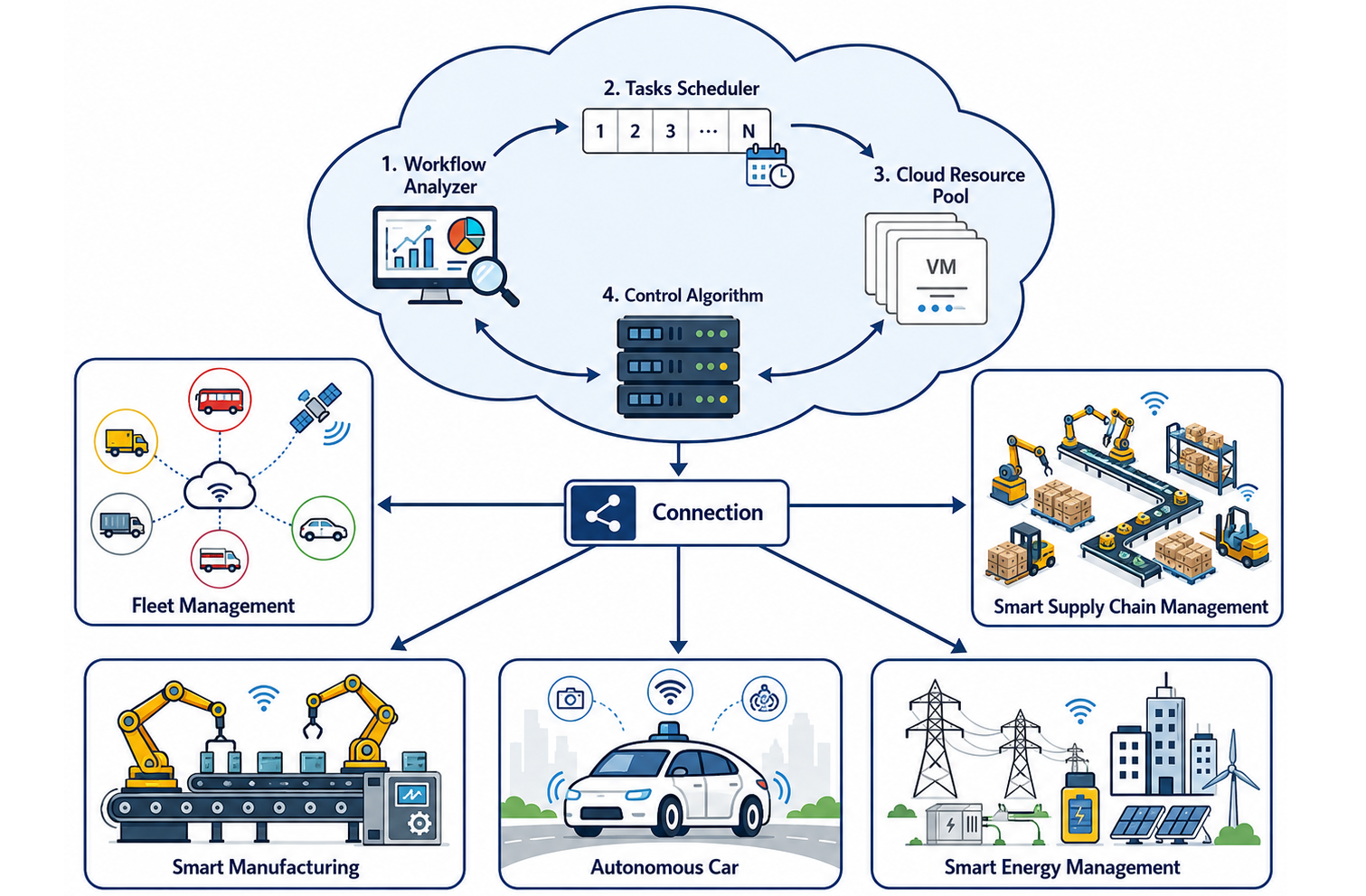}
\caption{Cloud-based on-demand networked control architecture. Local plants may request remote optimization or supervisory control services, but deployability depends on bounded communication delay, authenticated commands, and local fallback when the cloud service is late, unavailable, or rejected by a safety monitor.}\label{fig:on_demand_control}
\end{figure}
\subsection{Cloud-Assisted Privacy-Preserving Data-Driven Control}
A second major stream of work focuses on privacy-preserving data-driven control, where cloud resources learn, optimize, or evaluate control laws without directly exposing sensitive plant data. This problem is central to CCSs because industrial data may reveal production schedules, proprietary models, occupant behavior, grid conditions, or patient information. In contrast to ordinary cloud analytics, privacy-preserving control must protect data while preserving feedback performance and timing.

Fig.~\ref{fig:on_demand_control} shows a generic cloud-based on-demand
control-service architecture. The figure highlights the difference between
requesting remote optimization or supervisory decisions and depending on the
cloud for fast feedback. This distinction is central for deployable CCSs because
remote services must be combined with bounded communication delay,
authenticated commands, safety-monitor rejection logic, and local fallback when
the cloud service is late or unavailable.

One notable study by \citet{li2022secure} introduced the Weighted Access Control Scheme (SEWAC), a secure and efficient approach for cloud-assisted Industrial IoT (IIoT). SEWAC enables fine-grained, client-defined access structures based on weighted attributes, striking a balance between access-policy expressiveness and ciphertext length. Online key generation reduces computational costs, and outsourced decryption shifts part of the cryptographic burden to the cloud. This type of approach is important for CCS deployment because access control determines who can request, modify, or infer control-relevant data in multi-tenant cloud environments.

More broadly, privacy-preserving data-driven CCSs should be assessed along three coupled axes. The first is \emph{information exposure}: whether raw measurements, model parameters, constraints, event times, gradients, or learned policies can be inferred by a cloud provider or another participant. The second is \emph{closed-loop distortion}: whether encryption, coding, noise injection, quantization, or secure aggregation changes estimation accuracy, constraint satisfaction, or stability margins. The third is \emph{deadline feasibility}: whether the protected computation can be completed within the sampling period or decision horizon. A privacy mechanism that is strong in isolation may still be unsuitable for control if it delays actuation, hides diagnostically important residuals, or prevents timely fallback.

\begin{figure}[!htbp]
    \centering
    \includegraphics[width=\textwidth]{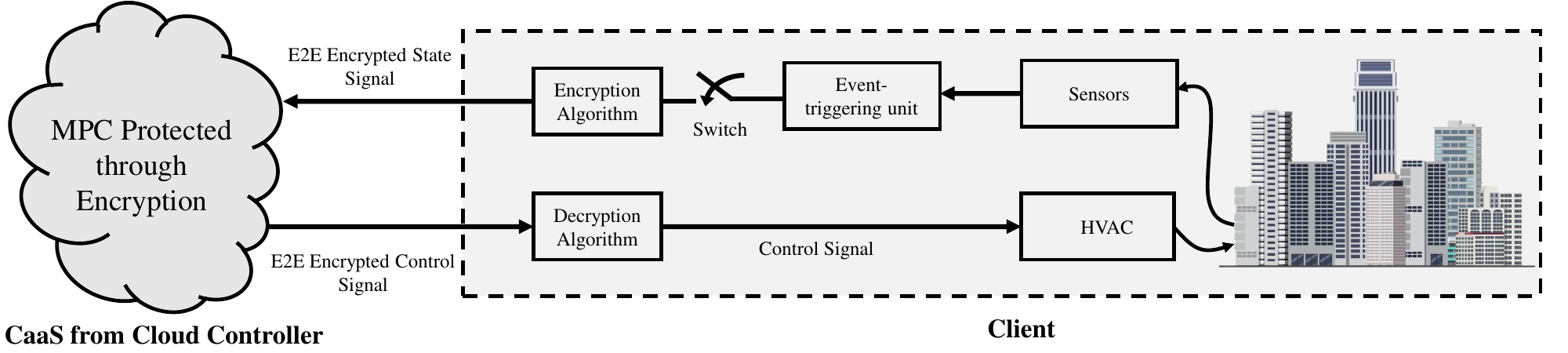}
\caption{Encrypted end-to-end communication between a cloud controller and client. The privacy value of the architecture depends not only on cryptographic strength, but also on whether encryption, key management, and outsourced computation preserve feedback deadlines, command integrity, and MPC feasibility.}
    \label{Fig11}
\end{figure}

\citet{PrivacyPreservingFramework2024ZhenanFeng} introduced an encrypted cloud-based framework for Heating, Ventilation, and Air-Conditioning (HVAC) control. The framework protects occupancy information while reducing communication and computation costs in encrypted event-based control. The proposed system, illustrated in Fig.~\ref{Fig11}, is significant because occupancy data is both privacy-sensitive and directly relevant to control decisions. Related work has used XOR-based methods, homomorphic encryption, dynamic coding-decoding, differential privacy, and probabilistic constraints to protect control data under uncertainty \citep{paper1, paper2, paper3, paper4, paper5}. These contributions show that privacy-preserving CCSs should be evaluated not only by cryptographic strength but also by control degradation, sampling-time feasibility, communication overhead, and leakage through event timing.
These research efforts offer innovative solutions for maintaining system functionality while safeguarding sensitive information. Table~\ref{tab:privacy_preserving_synthesis} summarizes representative privacy-preserving CCS approaches without assigning numerical scores. This choice is deliberate: the cited studies use different plants, sampling periods, cryptographic assumptions, hardware, and metrics, so a bar chart of normalized ratings would suggest a level of comparability that the literature does not yet support. The field still lacks common datasets, plant models, sampling periods, threat models, and reporting conventions for comparing privacy-preserving CCSs. Establishing such benchmarks is one of the most urgent requirements for making this literature cumulative.

Framework implication: privacy mechanisms should be part of the control design loop. A method that protects data but delays state estimation, hides safety-critical residuals, or prevents fallback validation is not yet a deployable privacy-preserving CCS.
\begin{table}[!htbp]
\centering
\caption{Privacy-preserving CCS mechanisms interpreted through closed-loop consequences}
\label{tab:privacy_preserving_synthesis}
\resizebox{0.95\textwidth}{!}{
\begin{tabular}{p{3.6cm}p{4.3cm}p{4.5cm}p{4.5cm}p{4.2cm}}
\toprule
\textbf{Approach} & \textbf{Protected information} & \textbf{Control-service value} & \textbf{Main closed-loop cost} & \textbf{Evidence caveat} \\
\midrule
\textbf{SEWAC access control} \citep{li2022secure} & Data access policies, ciphertexts, and client-defined attributes & Regulates who can request or infer IIoT control data in cloud-assisted settings & Key generation, policy management, and outsourced decryption overhead & Strong for access governance; less direct evidence for fast feedback timing \\
\midrule
\textbf{Encrypted event-based HVAC control} \citep{PrivacyPreservingFramework2024ZhenanFeng} & Occupancy and building-control information & Protects privacy-sensitive measurements while supporting cloud-assisted HVAC decisions & Encryption, event-triggering, and leakage through timing patterns must be accounted for & Evidence is application-specific; transfer to faster dynamics requires new timing studies \\
\midrule
\textbf{Encrypted LQG and quadratic optimization} \citep{alexandru2019encrypted, alexandru2020cloud} & States, costs, constraints, and multi-party optimization data & Connects formal privacy analysis to structured control computation & Homomorphic operations and transformations can constrain model class and timing & Mathematically strong for structured problems; nonlinear autonomy remains less mature \\
\midrule
\textbf{Encrypted ST-MPC and UAS MPC} \citep{naseri2022encrypted, feng2023secure} & Plant states, constraints, paths, and control policies & Enables cloud-side optimization while limiting information exposure & Key size, encoding, quantization, and solver overhead affect sampling feasibility & Reported results are not directly comparable across plants, hardware, and sampling periods \\
\midrule
\textbf{Blockchain-based detection} \citep{ramanan2021blockchain} & Shared evidence, attack reports, and replay-detection records & Supports integrity and post-event accountability across stakeholders & Consensus latency and storage overhead limit real-time feedback use & Best interpreted as audit support, not as a general privacy or fast-control mechanism \\
\bottomrule
\end{tabular}}
\end{table}
\subsection{Cloud-Based Privacy-Preserving LQG and Quadratic Optimization}
Cloud-based Linear Quadratic Gaussian (LQG) control and quadratic optimization occupy an important middle ground between classical control theory and secure cloud computation. They retain enough mathematical structure for stability, optimality, and privacy analysis, while still representing control tasks that can become computationally demanding in multi-agent or multi-stakeholder settings. For example, \citet{liu2020coordinated} investigated coordinated control challenges in networked multi-agent systems within a distributed cloud environment. Their work introduced a distributed cloud predictive control framework designed to improve regulation while mitigating transmission delays between cloud servers and agent nodes. The framework combines multi-step state prediction with coordination optimization, allowing future hidden states to be estimated from historical behaviors and interactions across spatially distributed agents.

Similarly, \citet{alexandru2020cloud} presented a cloud-based protocol for constrained quadratic optimization with private data from multiple parties. The protocol uses a projected gradient method on the Lagrange dual problem and incorporates partial homomorphic encryption to protect communication. \citet{alexandru2019encrypted} proposed transformation-based methods for enforcing data privacy while limiting computational overhead, with explicit analysis of confidentiality and privacy loss. These studies are valuable because they move beyond data encryption as an add-on and treat privacy as part of the optimization architecture itself.

The remaining limitation is scalability to nonlinear, hybrid, and tightly coupled systems. Most privacy-preserving optimization protocols still rely on structural assumptions that simplify encrypted computation. Future CCSs will require algorithms that preserve formal guarantees while tolerating model uncertainty, nonconvexity, and time-varying network conditions.

In the framework, LQG and quadratic optimization provide a mathematically tractable bridge between classical secure control and deployable CCSs, but the field must demonstrate how these guarantees hold under nonlinear autonomy, cloud timing variability, and multi-party trust boundaries.

\subsection{Learning-Enabled and Data-Driven CCSs}
Learning-enabled CCSs extend the cloud role from computation to model evolution. Cloud infrastructure can collect fleet data, train perception or prediction models, update digital twins, perform anomaly detection, and distribute improved policies to edge devices \citep{tuli2023ai, minerva2020digitaltwin, jin2025cloud}. In robotics and autonomous systems, this pipeline can support shared mapping, predictive maintenance, adaptive energy management, and multi-robot coordination \citep{kehoe2015cloudrobotics, wang2021cloud, li2024cooperative}. The control implication is substantial: a cloud-trained model can change the closed-loop map from observations to actions, even if the low-level controller remains local.

For this reason, learning-enabled CCSs should not be treated as ordinary cloud AI services. A learned estimator, scheduler, policy, or anomaly detector must be validated before it influences actuation. Key requirements include uncertainty quantification, distribution-shift detection, safe exploration boundaries, runtime monitors, safety shields, staged rollout, rollback, controller-version tracking, and post-deployment audit \citep{alshiekh2018shielding, brunke2022safelearning, ames2017cbf, chen2018hjreachability, hsu2024safetyfilter}. Federated learning, differential privacy, homomorphic encryption, and secure multi-party computation can reduce data exposure \citep{wang2020edge, alexandru2020secure, huong2021detecting}, but they can also change accuracy, communication load, and update latency. In CCSs, these effects must be reported as closed-loop consequences rather than solely as machine-learning metrics.

The unresolved issue is evidence maturity. Many learning-enabled CCS proposals show improved prediction, detection, or scheduling in simulation, but fewer demonstrate safe rollout under distribution shift, adversarial data, cloud outages, or rollback after a bad update. The field needs benchmark protocols that evaluate learned model updates alongside timing traces, fallback behavior, and safety violations.

Implication for the CCS framework: Cloud learning is deployable only when paired with edge/local assurance. The cloud may improve models and policies, but the physical system must retain a certified path to reject, contain, or roll back unsafe intelligence.

Table~\ref{tab:control_synthesis} summarizes the representative cloud-based control approaches reviewed in this section. The purpose of the table is not to rank methods by nominal performance, but to compare their cloud-side role, main benefit, primary limitation, and the evidence needed before stronger deployment claims can be made. This synthesis reinforces the central point that cloud-side computation becomes a control contribution only when timing, privacy, fallback, and safety evidence are reported together.

\begin{table}[!htbp]
\centering
\caption{Technical synthesis of representative cloud-based control approaches}
\label{tab:control_synthesis}
\resizebox{\textwidth}{!}{
\begin{tabular}{p{3.5cm}p{4.2cm}p{4.2cm}p{4.2cm}p{4.2cm}}
\toprule
\textbf{Approach} & \textbf{Cloud-side role} & \textbf{Main benefit} & \textbf{Primary limitation} & \textbf{Evidence needed for stronger deployment claims} \\
\midrule
\textbf{Cloud-based MPC} & Solves constrained optimization, prediction, and coordination tasks & Handles large horizons, nonlinear constraints, and fleet-level coordination & Sensitive to latency, jitter, and outage; real-time guarantees are difficult & End-to-end delay reporting, fallback logic, stability under network uncertainty \\
\midrule
\textbf{Privacy-preserving control} & Computes or assists control decisions using encrypted or protected data & Enables cloud services without exposing sensitive measurements or models & Encryption and privacy mechanisms add computation, communication, and possible performance loss & Common benchmarks, explicit threat models, and control-performance/privacy trade-off curves \\
\midrule
\textbf{Cloud LQG and quadratic optimization} & Executes structured estimation, control, and multi-party optimization & Allows formal analysis of privacy, optimality, and convergence & Often limited to linear, convex, or structurally simplified systems & Extension to nonlinear and coupled systems with verified sampling-time feasibility \\
\midrule
\textbf{Learning-enabled CCSs} & Trains models, updates policies, detects anomalies, and supports prediction & Exploits large-scale data and elastic computation for adaptive control & Guarantees for safety, interpretability, and robustness remain limited & Certified learning pipelines, uncertainty quantification, and safe online update mechanisms \\
\bottomrule
\end{tabular}}
\end{table}

\subsection{Workflow Scheduling, Orchestration, and Middleware}
The previous subsections treat the cloud as an optimizer or learner, but from a computer-science standpoint the more general object is a \emph{workflow}: a directed graph of sensing, estimation, optimization, learning, and actuation tasks whose placement and ordering must respect data dependencies and the decision deadline of Eq.~\eqref{eq:cloud_mpc_deadline}. Casting CCS execution as scheduling makes the systems character explicit, because each task inherits a release time, an execution-time distribution, a communication cost, and a layer assignment $\ell_k\in\{\mathrm{local},\mathrm{edge},\mathrm{fog},\mathrm{cloud}\}$. \citet{gao2026cloudenabled} make this concrete by formulating workflow-based real-time MPC for autonomous-vehicle trajectory tracking, where control tasks are scheduled as a cloud workflow under timing constraints. Deadline-constrained workflow scheduling results for cloud platforms more broadly show that execution-time allocation and task placement are first-class reliability concerns rather than implementation details \citep{yang2026continuous}. The control-relevant question is not only average makespan but whether the worst-case schedule still satisfies the per-step deadline with the required probability $\Pr(\tau_k\le T_s)\ge 1-\epsilon$.

Middleware determines whether such schedules are realizable. Robotic and industrial CCSs increasingly rely on publish-subscribe middleware such as ROS~2 and DDS, whose Quality-of-Service (QoS) policies govern discovery, reliability, deadline, liveliness, and security of message flows \citep{macenski2022ros2}. These QoS settings directly shape the communication terms in Eq.~\eqref{eq:cloud_mpc_deadline}, and misconfigured reliability or history policies can silently violate sampling deadlines even when the optimizer is fast. From the closed-loop service perspective, middleware QoS, orchestration, and workflow scheduling are therefore part of the correctness argument: a CCS is dependable only when task placement, message timing, and fallback switching are co-designed with the controller rather than delegated to a best-effort runtime.

\section{Security, Privacy, and Trust in the Feedback Loop}
\label{sec:security_trust}
Security and privacy in CCSs cannot be treated as conventional information security add-ons, because attacks may directly alter closed-loop behavior \citep{cardenas2008securecontrol, TEIXEIRA2015135}. Outsourcing sensing, estimation, optimization, or supervisory decisions to third-party cloud platforms creates multiple attack surfaces: physical communication channels, local sensors and actuators, cloud APIs, stored data, controller code, access policies, and cloud-side computation. Fig.~\ref{fig:cyber_resilience_map} first maps representative attack effects to control-aware defenses, while Fig.~\ref{fig:intrusion_points} illustrates where such attacks can enter a representative CCS architecture. A useful review of this literature must therefore ask three linked questions: what information is protected, what control function is exposed, and what stability or safety consequence follows if protection fails.

Research on secure CCSs can be grouped into confidentiality-preserving, integrity-preserving, and availability-preserving mechanisms \citep{cardenas2008securecontrol, TEIXEIRA2015135}. Encryption, homomorphic computation, secure multi-party computation, and differential privacy primarily protect confidentiality, but they can introduce delays and approximation errors that affect control performance \citep{naseri2022encrypted, feng2023secure}. Blockchain, secure logging, and distributed verification primarily support integrity and auditability, but may increase communication and consensus overhead \citep{ramanan2021blockchain}. Resilient control under DoS and FDI attacks addresses availability and correctness of feedback information, but often requires redundancy, observers, event-triggered communication, or fallback controllers \citep{yang2019predictive, yin2022cloud, schierman2020runtime}. The most mature CCS designs will need combinations of these mechanisms rather than isolated defenses.

At the control-loop level, common attack effects can be abstracted as
\begin{equation}
\vect{y}^{a}_k = \vect{y}_k + \vect{a}^{y}_k,\qquad
\vect{u}^{a}_k = \vect{u}_k + \vect{a}^{u}_k,\qquad
\gamma^{a}_k \in \{0,1\},
\label{eq:ccs_attack_model}
\end{equation}
where $\vect{a}^{y}_k$ denotes sensor or estimator corruption, $\vect{a}^{u}_k$ command manipulation, and $\gamma^{a}_k=0$ an unavailable or blocked communication/service event. Confidentiality mechanisms seek to prevent the inference of $\vect{x}_k$, $\vect{y}_k$, $\vect{u}_k$, model parameters, or policies; integrity mechanisms seek to detect or prevent nonzero $\vect{a}^{y}_k$ and $\vect{a}^{u}_k$; and availability mechanisms seek to maintain acceptable behavior when $\gamma^{a}_k=0$ for bounded or adversarial intervals.

Fig.~\ref{fig:cyber_resilience_map} maps these abstract attack effects to control-aware defenses. Sensor corruption, cloud-service unavailability, command tampering, and information leakage affect different parts of the feedback loop, so defenses must be evaluated based on their effects on stability, safety, delay, privacy, and recovery under degraded operation.

\begin{figure}[!t]
\centering
\resizebox{0.98\textwidth}{!}{%
\begin{tikzpicture}[
font=\footnotesize,
>=Latex,
line cap=round,
line join=round,
core/.style={draw=black!65, fill=gray!7, rounded corners=2pt, align=center, text width=2.65cm, minimum height=0.86cm},
attack/.style={draw=red!70!black, fill=red!7, rounded corners=2pt, align=center, text width=2.65cm, minimum height=0.76cm},
defense/.style={draw=green!45!black, fill=green!8, rounded corners=2pt, align=center, text width=2.85cm, minimum height=0.76cm},
flow/.style={-Latex, thick, black!60},
bad/.style={-Latex, thick, dashed, red!75!black},
good/.style={-Latex, thick, green!45!black}
]
\node[attack] (fdi) at (-5.85,2.0) {False data\\injection};
\node[attack] (dos) at (-1.95,2.0) {DoS/cloud\\outage};
\node[attack] (tamper) at (1.95,2.0) {Command\\tampering};
\node[attack] (leak) at (5.85,2.0) {Eavesdropping\\model leakage};
\node[core] (sensor) at (-5.85,0) {Sensors and\\estimators};
\node[core] (cloud) at (-1.95,0) {Cloud/edge\\controller};
\node[core] (actuator) at (1.95,0) {Actuators and\\local plant};
\node[core, fill=blue!6, draw=blue!55!black] (objective) at (5.85,0) {System stability\\ and privacy};
\node[defense] (observer) at (-5.85,-2.0) {Secure observer\\residual tests};
\node[defense] (fallback) at (-1.95,-2.0) {Predictive buffer\\fallback control};
\node[defense] (auth) at (1.95,-2.0) {Command\\ validation};
\node[defense] (privacy) at (5.85,-2.0) {Encrypted\\ computation};
\draw[flow] (sensor) -- node[above, font=\scriptsize] {$\vect{y}_k$} (cloud);
\draw[flow] (cloud) -- node[above, font=\scriptsize] {$\vect{u}_k$} (actuator);
\draw[flow] (actuator) -- (objective);
\draw[bad] (fdi.south) -- node[right, font=\scriptsize, text=red!70!black] {$\vect{a}^{y}_k$} (sensor.north);
\draw[bad] (dos.south) -- node[right, font=\scriptsize, text=red!70!black] {$\gamma_k^a=0$} (cloud.north);
\draw[bad] (tamper.south) -- node[right, font=\scriptsize, text=red!70!black] {$\vect{a}^{u}_k$} (actuator.north);
\draw[bad] (leak.south) -- node[right, font=\scriptsize, text=red!70!black] {leakage} (objective.north);

\draw[good] (observer.north) -- (sensor.south);
\draw[good] (fallback.north) -- (cloud.south);
\draw[good] (auth.north) -- (actuator.south);
\draw[good] (privacy.north) -- (objective.south);
\draw[good] (observer.east) -- (fallback.west);
\draw[good] (fallback.east) -- (auth.west);
\draw[good] (auth.east) -- (privacy.west);

\node[draw=black!50, fill=yellow!12, rounded corners=2pt, align=center, font=\scriptsize, text width=13.4cm, inner sep=2pt] at (0,-3.18) {Assess defenses by stability, safety, delay, privacy, and recovery under degraded operation.};
\end{tikzpicture}%
}
\caption{Cyber-resilience map for CCSs. Sensor corruption, service unavailability, command manipulation, and information leakage affect different parts of the feedback loop. Resilient CCS design maps each threat to control-aware defenses such as secure observers, predictive buffers, fallback controllers, command authentication and validation, encrypted computation, and privacy-preserving data handling.}
\label{fig:cyber_resilience_map}
\end{figure}
\begin{figure}[!htbp]
    \centering
    \includegraphics[width=\textwidth]{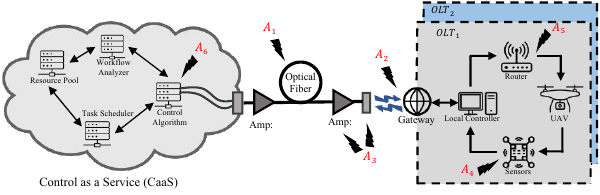}
    \caption{Representative intrusion points in an optical-fiber CCS architecture. Physical-layer attacks, local host-network attacks, feedback-loop manipulation, and cloud-resource compromise map to different control consequences, so the appropriate defense depends on whether telemetry, commands, controller code, or cloud computation is exposed.}
    \label{fig:intrusion_points}
\end{figure}
\subsection{Control-Aware Security Objectives}
For CCSs, confidentiality, integrity, availability, and accountability must be interpreted in closed-loop terms. Confidentiality protects states, inputs, constraints, costs, learned policies, digital-twin parameters, and event times from unauthorized inference. It is addressed through encrypted communication, homomorphic encryption, secure multi-party computation, differential privacy, and careful data minimization. The control cost is delay, quantization, approximation, or loss of diagnostic visibility, so privacy claims should be accompanied by timing and performance evidence.

Integrity protects the correctness of measurements, model updates, optimization results, controller binaries, and actuator commands. Authentication, command validation, secure observers, FDI detection, replay detection, signed controller artifacts, and secure logs are relevant only if they prevent or expose changes that matter for stability and safety \citep{cardenas2008securecontrol, mo2009replay, AttackDetection2013, TEIXEIRA2015135}. A cryptographically authenticated command may still be unsafe if it is stale, computed from corrupted states, or issued by an authorized but compromised cloud service.

Availability protects the plant's ability to keep operating acceptably when packets are dropped, channels are jammed, a cloud region fails, or an adversary induces resource exhaustion. DoS-resilient control, predictive buffers, redundant edge/fog services, local fallback, and graceful degradation are therefore availability mechanisms with direct control content \citep{yang2019predictive, yin2022cloud, schierman2020runtime}. Finally, accountability concerns the ability to reconstruct what controller version, model update, cloud request, human approval, or security event contributed to a physical outcome. Audit trails and blockchain-style ledgers are most useful for multi-stakeholder evidence and post-event accountability, not for fast feedback loops.

Framework implication: a defense is mature only when its protection target, location in the loop, control function, overhead, and deployment evidence are explicit. Security-by-design in CCSs means co-designing cryptography, detection, fallback, and governance with feedback timing and safety constraints.

\subsection{Encrypted Cloud-Based MPC}
The combination of cryptographic techniques and control-aware algorithm design is crucial for secure online control. In encrypted control, third-party cloud platforms can evaluate controller operations using ciphered or partially shared data without full access to plant parameters, state variables, or control policies. \citet{naseri2022encrypted} proposed an encrypted Set-Theoretic Model Predictive Control (ST-MPC) framework for cloud-based NCSs. Their approach integrates a modified ST-MPC scheme with an additive homomorphic cryptosystem, enabling secure execution on encrypted data while preserving confidentiality. To maintain real-time applicability, the method incorporates zonotopic inner approximations and a half-space projection procedure that reduces the computational burden on the smart actuator. The two-tank simulation demonstrates that encrypted cloud control can preserve both privacy and closed-loop objectives when the optimization structure is carefully adapted to the cryptographic scheme.

Building on this line of research, \citet{feng2023secure} addressed privacy challenges in path-following Unmanned Aerial Systems (UAS) operating under adverse communication conditions. To protect against potential cyber threats and privacy breaches, they proposed a secure MPC framework that integrates encrypted MPC with an efficient encoding/decoding scheme based on Paillier cryptography. This framework explicitly accounts for adversarial threats in both remote control stations and communication links. Under the assumptions studied, encrypted MPC can support closed-loop stability while protecting selected UAS information. The deployment lesson is cautious: operational security and data confidentiality must be evaluated together with timing, network, and fallback evidence in safety-critical applications.

Table~\ref{tab:encrypted_mpc_synthesis} compares encrypted ST-MPC and secure UAS MPC at the level supported by the reported evidence. The two studies use different plant models, cryptographic configurations, computational settings, and performance metrics, so strict numerical normalization would be misleading without a shared benchmark. The important synthesis is qualitative: encrypted MPC can protect states, constraints, paths, and policies, but its control value depends on whether cryptographic overhead is compatible with the relevant sampling or decision deadline.
\begin{table}[!htbp]
\centering
\caption{Encrypted MPC studies interpreted as control-service evidence rather than strict numerical benchmarks}
\label{tab:encrypted_mpc_synthesis}
\resizebox{0.95\textwidth}{!}{
\begin{tabular}{p{3.8cm}p{4.4cm}p{4.5cm}p{4.5cm}p{4.2cm}}
\toprule
\textbf{Study} & \textbf{Control task} & \textbf{Privacy/security mechanism} & \textbf{Real-time implication} & \textbf{Comparison caveat} \\
\midrule
\citet{naseri2022encrypted} & Set-theoretic MPC for a cloud-based NCS & Additive homomorphic cryptosystem with control-oriented approximations & Emphasizes predictable actuator-side computation and feasibility under the studied model & Evidence is tied to the reported plant, model structure, and cryptographic implementation \\
\midrule
\citet{feng2023secure} & Secure MPC-based UAS path following under adverse communication & Paillier-based encryption with encoding/decoding for protected remote control & Shows that key size and encrypted computation affect per-step timing and path-following feasibility & Path scenarios, hardware, key settings, and metrics differ from ST-MPC studies \\
\midrule
\textbf{Cross-study lesson} & Cloud-assisted optimization with confidentiality constraints & Privacy-preserving computation inside or near the feedback loop & Encryption strength is meaningful for control only when paired with deadline, stability, and fallback evidence & A shared benchmark should report plant, sampling period, hardware, network traces, key sizes, and safety metrics \\
\bottomrule
\end{tabular}}
\end{table}
\subsection{Encryption in Distributed Control and Optimization}
Distributed CCSs introduce a second privacy problem: agents may need to coordinate without revealing their true states, constraints, costs, or local objectives to neighbors or cloud coordinators. This is different from protecting a single client against an external cloud host. It requires privacy among cooperating entities as well as against external adversaries. Encrypted distributed optimization, secure aggregation, and transformation-based privacy mechanisms can support cooperation, but they must be integrated with consensus, estimation, and constraint-handling algorithms. In this regard, \citet{feng2023secure} addressed distributed UAS security through a secure MPC framework for path-following under adverse network conditions. The broader implication is that distributed encrypted control should be evaluated by both cryptographic privacy and coordination quality: convergence speed, constraint satisfaction, communication load, and robustness to compromised agents.
\subsection{Auditability through Blockchains}
Blockchain technology, a decentralized ledger maintained through Peer-to-Peer (P2P) networks and secured by cryptographic methods, has attracted interest for strengthening integrity and auditability in cyber-physical control systems. In CCSs, blockchain is most relevant when multiple stakeholders must share control-relevant evidence without trusting a single central party. In large-scale power systems, for example, regional utilities continuously generate sensor data, and coordinated replay attacks can exploit limitations in centralized data sharing. \citet{ramanan2021blockchain} proposed a decentralized blockchain-based architecture for detecting coordinated replay attacks while preserving sensor information integrity. Their method uses Bayesian inference with locally detected attack probabilities configured within a blockchain framework and was evaluated against a broadcast-gossip-driven distributed algorithm on a private Ethereum blockchain. The value of this line of work lies in tamper resistance and shared accountability; its limitations lie in latency, consensus costs, and the difficulty of integrating ledger mechanisms into fast feedback loops.

\subsection{Security and Stability in the Face of DoS Attacks}
DoS attacks are especially damaging to CCSs because they target the availability of communication and cloud services on which the feedback loop may depend \citep{L-yasirstabilizing}. A short outage may be tolerable for slow supervisory control, but catastrophic for fast dynamics if no local fallback exists. \citet{yin2022cloud} proposed a cloud-based compensation control method that combines adaptive event-triggering and predictive control to alleviate DoS and delay effects. \citet{li2024cooperative} proposed a cloud-fog collaborative framework for heterogeneous swarms, combining fog-level distributed consensus with cloud-level leader-following control to sustain cooperation under DoS disruptions. \citet{yin2024secure} employed reinforcement learning in an event-triggered framework to counteract bandwidth limitations and DoS attacks, while \citet{guo2024cloud} studied cloud-edge MPC under cyber attacks. Recent resilient MPC work further sharpens the availability problem by treating DoS and hybrid attacks within tube-based predictive-control designs \citep{yang2026resilientoutput, shen2026resilienttube}. Collectively, these contributions show a shift toward multi-layered resilient architectures in which local, fog, and cloud resources share responsibility for maintaining stability during degraded communication. A central open issue is how to certify fallback transitions and recovery behavior when cloud services reappear after an attack.
\subsection{Resilient Control under False Data Injections}
In CCSs, resilience refers to maintaining stability and performance in the face of both accidental faults and malicious cyber-physical attacks. False Data Injection (FDI) attacks, in which adversaries alter sensor, estimator, or control data, are particularly serious because they can cause the cloud controller to optimize based on a false representation of the plant. Over the past decade, FDI mitigation has evolved from model-based residual analysis \citep{AttackDetection2013, NetworkInducedConstraints2013} toward observer-based, statistical, data-driven, and distributed detection \citep{GameTheoretic2016, TEIXEIRA2015135, 9406118, 8027127}. Recent work has also examined cooperative security analysis of industrial cloud control systems under FDI attacks \citep{zhao2024cooperative} and vulnerability analysis for nonlinear state estimation under joint deception attacks \citep{li2026vulnerability}. Privacy-preserving distributed schemes now allow collaborative detection without sharing raw measurements \citep{paper2, paper3}, and integration with resilient control enables reconfiguration or switching to trusted edge computation. The unresolved challenge is stealth: advanced attackers may inject signals that remain statistically plausible while still biasing optimization, so FDI defense must be connected to control objectives rather than evaluated only as anomaly detection.

Table~\ref{tab:security_synthesis} maps common CCS security mechanisms to control-loop objectives. This mapping clarifies that encrypted control, differential privacy, blockchain logging, DoS-resilient control, and FDI detection protect different parts of the feedback loop and introduce different costs for latency, computation, communication, and certification. The table also helps distinguish mechanisms that protect confidentiality from those that support integrity, availability, resilience, or accountability.

\begin{table}[!htbp]
\centering
\caption{Security mechanisms in CCSs mapped to control-loop objectives}
\label{tab:security_synthesis}
\resizebox{\textwidth}{!}{
\begin{tabular}{p{3.8cm}p{4.2cm}p{4.2cm}p{4.2cm}p{4.2cm}}
\toprule
\textbf{Mechanism} & \textbf{Primary protection} & \textbf{Control-loop role} & \textbf{Main cost} & \textbf{Best-fit CCS setting} \\
\midrule
\textbf{Encrypted control} & Confidentiality of states, inputs, models, and policies & Allows cloud-side computation without exposing sensitive plant information & Cryptographic computation and communication overhead & Slow or medium-rate control where privacy is critical \\
\midrule
\textbf{Differential privacy} & Protection against inference from shared data or model updates & Supports data-driven control and learning across clients & Noise-induced accuracy and performance degradation & Analytics, learning, and aggregate supervision rather than fast inner loops \\
\midrule
\textbf{Blockchain and secure logging} & Integrity, auditability, and nonrepudiation & Records sensor evidence, attack reports, and multi-stakeholder control events & Consensus latency and storage overhead & Multi-operator infrastructure, energy systems, and post-event accountability \\
\midrule
\textbf{DoS-resilient control} & Availability under communication or service disruption & Maintains stability during missing packets, blocked channels, or cloud outage & Redundancy, prediction, and fallback-controller complexity & Safety-critical systems requiring graceful degradation \\
\midrule
\textbf{FDI detection and resilient estimation} & Integrity of measurements and control commands & Prevents optimization on falsified plant states & Observer, model, and data-quality requirements & Systems with high-value sensors and adversarial data channels \\
\bottomrule
\end{tabular}}
\end{table}

Table~\ref{tab:assurance_landscape} broadens this mapping into a CCS assurance taxonomy. It highlights formal verification and runtime assurance, encryption and homomorphic computation, differential privacy and secure aggregation, blockchain and secure logging, and quantum-safe or post-quantum communication mechanisms. These technologies should not be read as interchangeable alternatives. Each protects a different layer of the CCS stack, and each introduces different costs for latency, communication, computation, and certification. Quantum key distribution and post-quantum cryptography are best viewed as high-assurance infrastructure topics for long-lived CCS deployments; current evidence does not justify treating them as near-term solutions for ordinary robotic feedback loops.

\begin{table}[!htbp]
\centering
\caption{Assurance and security technologies as a CCS taxonomy rather than a one-size-fits-all solution}
\label{tab:assurance_landscape}
\resizebox{0.97\textwidth}{!}{
\begin{tabular}{p{3.6cm}p{4.3cm}p{4.3cm}p{4.3cm}p{4.3cm}}
\toprule
\textbf{Technology family} & \textbf{Primary CCS value} & \textbf{Where it belongs in the loop} & \textbf{Main limitation} & \textbf{Best deployment interpretation} \\
\midrule
\textbf{Formal verification and runtime assurance} & Stability, safety, and policy compliance evidence & Local/edge safety monitors, controller verification, safety cases & Model abstraction gap and difficulty covering learned or cloud-updated behavior & Essential for fallback, switching, and safety-critical autonomy \\
\midrule
\textbf{Encryption and homomorphic computation} & Confidentiality of states, models, constraints, and control inputs & Communication links, cloud optimization, privacy-preserving MPC & Computation, communication, quantization, and latency overhead & Useful when privacy is central, and deadlines are explicitly validated \\
\midrule
\textbf{Differential privacy and secure aggregation} & Limits inference from shared data, model updates, and fleet telemetry & Learning, analytics, aggregate estimation, federated training & Noise or aggregation can degrade estimation and control performance & Better suited to learning and supervision than fast inner loops \\
\midrule
\textbf{Blockchain and secure logging} & Shared integrity, nonrepudiation, and post-event accountability & Audit trails, multi-stakeholder evidence, incident reconstruction & Consensus latency, storage burden, and weak fit to fast feedback & Useful for governance and evidence, not as a general real-time control defense \\
\midrule
\textbf{QKD and post-quantum cryptography} & Long-lived high-assurance channel security & Backbone links, supervisory services, key-management infrastructure & Specialized infrastructure, maturity, and uncertain control-loop benefit & Future-facing infrastructure topic for high-value CCSs rather than routine robotic feedback \\
\bottomrule
\end{tabular}}
\end{table}
\section{Deployment Evidence and Cloud-Control Platforms}
\label{sec:deployment_maturity}
Industrial automation is a decisive test case for CCSs because it combines large-scale sensing, stringent reliability requirements, safety constraints, proprietary data, and pressure to optimize costs. In this setting, the value of cloud control lies beyond mere remote computation. It is the ability to deliver control, monitoring, diagnosis, learning, and optimization as continuously maintained cyber-physical services across fleets of assets and geographically distributed plants \citep{alavanthan2023cloud, awotunde2022big, LEI2016324, gao2023predictive}. The industrial literature can be read along two complementary lines: service-oriented cloud control, where selected functions are exposed as scalable cloud services, and platform-oriented CCS design, where communication, data management, security, middleware, operator interfaces, and deployment workflows are integrated into an operational architecture.
\subsection{Cloud Control as a Service for Industrial Automation}
Cloud Control as a Service decouples selected control functions from fixed local hardware and makes them available through virtualized, remotely managed resources \citep{givehchi2014control, wang2021cloud}. This model can improve flexibility, reduce duplication of specialized computing infrastructure, and support continuous monitoring, predictive maintenance, and long-horizon optimization \citep{xia2024cloud, 13-xia2012networked}. Examples include grain storage monitoring via cloud-connected IoT devices \citep{alavanthan2023cloud} and secure access control for cloud-assisted IIoT via SEWAC \citep{li2022secure}. For industrial adoption, however, the service model must define responsibility boundaries: which control tasks remain local, what happens during cloud outage, how controller updates are approved, and how performance guarantees are verified after deployment.
\subsection{Recent Platform Designs of CCSs for Industrial Automation}
Platform-oriented CCS designs integrate cloud infrastructure, communication protocols, data pipelines, security mechanisms, and operator interfaces. \citet{awotunde2022big} proposed an IoT-cloud framework on the Cloudera platform for continuous healthcare monitoring, while \citet{wang2021cloud} introduced a cloud-based mission-control architecture for Unmanned Surface Vehicles validated through sea trials. \citet{gao2023predictive} demonstrated cloud-based predictive cruise control for reducing energy consumption and traffic delays, and \citet{yin2022cloud} addressed DoS and delay mitigation in multi-agent systems. Privacy-preserving optimization and secure MPC designs \citep{alexandru2020cloud, feng2023secure} further show how cloud platforms can support safety-critical and adversarial environments.

The main lesson is that industrial CCS maturity should be judged by the quality of evidence,. Simulation results are useful, but a high-level synthesis should distinguish simulation, hardware-in-the-loop testbeds, prototype cloud services, field trials, and certified deployments. Future industrial CCS platforms will need standardized timing reports, failure-mode documentation, fallback-controller validation, secure update pipelines, operator workflow analysis, cloud-provider dependency assessment, and transparent cost models. Without this evidence, claims of scalability and security remain promising but incomplete.

\subsection{Standards, Certification, and Regulatory Compliance}
Trustworthy-CCS claims ultimately rest on recognized assurance frameworks, yet the literature rarely connects to them explicitly. Functional-safety standards such as IEC~61508 and its automotive derivative ISO~26262 define safety-integrity levels and hazard-analysis processes that a certified fallback controller must satisfy, while DO-178C plays the analogous role for airborne autonomy. Industrial communication and control security is governed by IEC~62443 (formerly ISA-99) for industrial automation and control systems, and IEC~61499 addresses the portability of distributed control software that is directly relevant to cloud-edge orchestration. On the computing side, ISO/IEC~27001 and the cloud-specific ISO/IEC~27017 and ISO/IEC~27018 frame information-security management and tenant-data governance, complementing the data-protection regimes (GDPR, CCPA) noted elsewhere in this review \citep{qi2020efficient, froomkin2022safety}. The open problem is that none of these standards was written for cloud-in-the-loop feedback: mapping CCS fallback, update, and audit evidence onto safety-integrity levels and IEC~62443 security levels, and certifying the transitions between cloud, fog, edge, and local authority, remain unsolved but necessary steps toward certifiable deployment.

For deployment, the framework implication is direct: maturity is not a final appendix to the control problem. It determines what claims can responsibly be made. A simulation can support algorithmic feasibility; a testbed can support timing and integration claims; a field trial can support operational feasibility; and a certified service requires documented fallback, update, audit, and recovery processes.

\section{Operational Trust, Governance, and Data Protection}
\label{sec:operational_trust}
Centralized cloud control creates an operational trust problem: the cloud platform is not only a data processor but also part of the feedback path. Security controls must therefore be designed around control-service continuity, command integrity, model confidentiality, auditability, and accountable recovery. This section reframes conventional data-protection measures as operational requirements for CCSs rather than as generic cybersecurity practices.
\subsection{Data Confidentiality and Key Management}
Encryption remains a foundational requirement for CCSs, but its role differs across the control stack. Data in transit must be protected against interception and manipulation, stored data must be protected against cloud-side compromise, and control computations may require privacy-preserving mechanisms when the cloud is not fully trusted. Standard mechanisms such as Advanced Encryption Standard (AES) \citep{L-alemami2023advanced} and Rivest-Shamir-Adleman (RSA) \citep{L-liu2023rivest} are useful for communication and storage, while chaotic internal encryption and coding-based schemes may provide additional physical-layer protection in networked CCSs \citep{2-ali2023chaos, L-yasiraccess, yasiralicybernetics}. The control question is not only which cipher is strong, but where encryption and decryption occur relative to sensing, estimation, optimization, and actuation. Encryption delay enters the deadline in Eq.~\eqref{eq:cloud_mpc_deadline} through uplink processing, cloud-side computation, downlink processing, and actuator-side application. A security mechanism that protects telemetry but causes missed sampling deadlines can degrade closed-loop safety.

Key management is similarly control-relevant. A compromised telemetry key can expose plant states, operating constraints, proprietary models, or user behavior; a compromised command-signing key can allow unauthorized actuation; and a compromised update key can distribute an unsafe controller or model. CCS deployments therefore need key rotation, revocation, secure storage, and incident procedures that specify whether the plant should continue under degraded local control, reject cloud updates, or trigger fallback after a key event. Key-management evidence should include not only cryptographic strength but also recovery time, controller-version integrity, and the effect of rekeying on service availability.
\subsection{Access Control for Cloud-Hosted Control Services}
Access control in CCSs must regulate who can read plant data, change model parameters, deploy controllers, request cloud optimization, issue actuator commands, trigger fallback, or approve recovery after an incident. Role-Based Access Control (RBAC) and least-privilege policies \citep{L-RBAC} provide a baseline, while Multi-Factor Authentication (MFA) reduces credential-compromise risk \citep{otta2023systematic}. For control services, however, roles should be tied to physical consequences. A monitoring user may view telemetry but not issue commands; a controller engineer may upload a candidate policy but not approve deployment; an operator may trigger an emergency fallback but not modify cloud-optimization code. Access logs should therefore preserve controller-version, model-update, approval, and command histories that can support post-event analysis.

Firewalls, Intrusion Detection Systems (IDS), VPN-based remote access, and cloud API controls remain relevant \citep{L-pradhan2020intrusion, hyder2023toward}, but their value must be stated in loop terms. They may protect telemetry streams, command channels, model-update paths, operator sessions, and cloud optimization endpoints. They do not by themselves guarantee safe control: an authenticated and encrypted command can still be stale, infeasible, or unsafe if generated from corrupted data. Secure key exchange, for example through Elliptic Curve Diffie-Hellman (ECDH), helps establish protected channels between local controllers and cloud services \citep{tanksale2024efficient}, but command validation and runtime monitors are still needed near the plant.
\subsection{Quantum-Safe and Physical-Layer Key Exchange}
Quantum communication and post-quantum cryptography are relevant to CCSs mainly because long-lived industrial infrastructure may remain in service beyond the useful lifetime of current public-key assumptions. Quantum Key Distribution (QKD) can support secure key exchange by making eavesdropping detectable through quantum-state disturbance \citep{L-kong2024secret}. In practical CCS deployments, QKD is more likely to secure high-value backbone links, supervisory channels, or key-management infrastructure than fast local robotic feedback loops, because it requires specialized transmitters, receivers, and compatible networking infrastructure. Hybrid approaches that combine QKD or post-quantum key establishment with classical symmetric encryption may support high-assurance cloud control environments \citep{ricci2024hybrid}. The current evidence does not justify presenting QKD or post-quantum methods as near-term solutions for ordinary feedback timing; they are better treated as future-facing infrastructure for high-value CCS deployments.
\subsection{Monitoring, Auditing, and Runtime Assurance}
Continuous monitoring and auditability are essential because CCS failures can arise from software bugs, cloud misconfiguration, network anomalies, insider misuse, or cyber-physical attacks. Security audits verify policy compliance and configuration integrity, while runtime monitoring tracks control commands, cloud requests, timing deviations, anomaly scores, and fallback-controller activations \citep{L-torkura2021continuous}. For CCSs, monitoring should be connected to control semantics: a suspicious event is not only a login anomaly or packet spike, but also a deviation in closed-loop timing, command plausibility, estimator residuals, constraint satisfaction, or model-update behavior.

Audit mechanisms should record the artifacts needed to reconstruct a physical outcome: controller version, model parameters, digital-twin state, optimization request, solver result, cloud service endpoint, human approval, security alert, fallback trigger, and recovery action. Blockchain or distributed ledgers may help when multiple organizations need shared integrity and nonrepudiation, but they should not be placed in fast feedback paths unless latency and storage overhead are irrelevant to the control decision \citep{ramanan2021blockchain}. Their best role is post-event accountability, evidence sharing, and safety-case support.
\subsection{Recovery, Patch Management, and Human Factors}
Backup and recovery plans must protect both information and control continuity. Recoverable CCS artifacts include controller versions, model parameters, digital-twin states, fallback policies, audit logs, safety cases, cryptographic keys, configuration files, and deployment manifests. These artifacts should be backed up in geographically and logically separated locations, with recovery procedures tested under cloud outage and attack scenarios \citep{L-li2019dynamic}. A recovery plan that restores data but not a verified controller, fallback policy, or key hierarchy is incomplete from a control-service perspective.

Patch management is equally important because cloud-hosted controllers depend on operating systems, container images, communication libraries, middleware, and optimization solvers that may carry vulnerabilities \citep{L-dissanayake2022and}. A patch may change solver timing, numerical behavior, dependency versions, communication latency, or the attack surface of a controller service. Recent deadline-constrained workflow scheduling results for cloud systems reinforce why execution-time allocation is part of the control-service evidence base rather than a generic IT concern \citep{yang2026continuous}. CCS patch pipelines should therefore include staged rollout, regression tests on representative plant models, deadline checks, rollback mechanisms, and explicit approval before safety-relevant updates reach production systems.

Human factors remain a major risk and a major defense. Operators and engineers need training in phishing resistance, secure key handling, controller-deployment approval, emergency override, fallback activation, and incident escalation \citep{L-thron2024human, L-martins2019specialized}. In human-facing autonomy, the operator workflow should specify who can pause cloud updates, reject unsafe commands, authorize degraded operation, and communicate with affected users or infrastructure owners. A mature CCS security program combines technical defenses with operational discipline.

Implication for the CCS framework: trustworthy cloud control requires a governance layer. Responsibility boundaries among the plant owner, cloud provider, controller designer, software maintainer, and operator should be visible in controller-version records, approval workflows, incident logs, and safety-case evidence.

\section{Autonomous and Networked Application Domains}
\label{sec:application_domains}
Autonomous and networked application domains are the practical proving ground for CCSs because they contain distributed sensors, heterogeneous communication links, safety constraints, and control objectives that span local response and system-level optimization. The following domains are not a generic IoT catalogue. They illustrate where cloud-hosted intelligence can support autonomy, coordination, and supervisory control, and they reveal why hybrid architecture, privacy protection, fallback certification, and operational accountability are essential. Table~\ref{tab:domain_comparison} summarizes the domains along four CCS-relevant axes: which functions are cloud-suitable, which must remain local or edge-resident, the dominant deployment constraint, and the evidence level the literature currently reaches.

\begin{table}[!htbp]
\centering
\caption{Autonomous application domains viewed through the CCS placement-and-evidence lens}
\label{tab:domain_comparison}
\begingroup
\footnotesize
\setlength{\tabcolsep}{4pt}
\renewcommand{\arraystretch}{1.05}
\resizebox{\textwidth}{!}{%
\begin{tabular}{p{2.6cm}p{3.6cm}p{3.4cm}p{3.2cm}p{2.6cm}}
\toprule
\textbf{Domain} & \textbf{Cloud/fog-suitable functions} & \textbf{Local/edge-resident functions} & \textbf{Dominant constraint} & \textbf{Typical evidence today} \\
\midrule
\textbf{Robotics and fleets} & Shared mapping, fleet scheduling, long-horizon planning, perception-model updates, digital twins & Stabilization, collision avoidance, force/torque loops, emergency stop & Hard real-time safety and middleware QoS & Prototype/testbed; few field trials \\
\midrule
\textbf{Transportation} & Predictive routing, signal coordination, mixed-traffic and platoon optimization & Vehicle stability, braking, fast collision avoidance & Latency under mobility and safety & Simulation; emerging field studies \\
\midrule
\textbf{Buildings} & Comfort and energy optimization, occupancy-aware scheduling, remote maintenance & Local comfort/safety overrides & Occupancy and behavioral privacy & Simulation and pilot deployments \\
\midrule
\textbf{Healthcare} & Analytics, anomaly detection, decision support, semi-closed-loop care & Device-level safety interlocks & Medical-data privacy and regulation & Monitoring studies; limited closed-loop \\
\midrule
\textbf{Smart cities} & City-scale prediction, multi-agency resource allocation, infrastructure monitoring & Local intersection/sensor control & Cross-operator governance and accountability & Conceptual and pilot scale \\
\midrule
\textbf{Smart grids} & Long-horizon optimization, demand forecasting, fleet-level DER analytics & Fast protection and local stability & Resilience under loss or attack & Simulation and testbed; some field \\
\bottomrule
\end{tabular}}
\endgroup
\end{table}

\subsection{Robotics, Multi-Robot, and Fleet-Level Autonomy}
Robotics is one of the clearest motivating domains for CCSs because robot autonomy naturally separates into functions with different timing and evidence requirements. Inner-loop stabilization, collision avoidance, force/torque control, and emergency stops require local or edge authority. By contrast, cloud services can support shared mapping, fleet scheduling, long-horizon planning, perception-model updates, simulation-based validation, predictive maintenance, and digital-twin synchronization. This layered view is especially relevant to UAV swarms, warehouse robots, mobile manipulation, unmanned surface vehicles, industrial robot cells, and human-facing service robots \citep{kehoe2015cloudrobotics, wang2021cloud, liu2017predictive, feng2023secure, li2024cooperative, jin2025cloud}. Safety-critical robotic operation also needs certified local constraints and fallback logic, a point reinforced by recent work on robust safety-critical control for robotic systems with time-varying constraints and input disturbances \citep{xiong2026robustsafety}.

Cloud robotics and multi-robot autonomy also expose the trust problem sharply. A cloud planner may generate a useful global route, but a local robot must be able to reject commands that violate safety envelopes, arrive too late, conflict with local perception, or were computed from stale maps. Middleware choices, such as ROS 2 and Data Distribution Service (DDS)-style communication, influence Quality of Service (QoS), discovery, message timing, and security boundaries; however, middleware support is not by itself a control guarantee \citep{macenski2022ros2}. CCS research for robotics should therefore report not only task performance but also message timing, missed-deadline behavior, safety-shield activation, rollback after model updates, and interaction with human operators \citep{schierman2020runtime, ames2017cbf, chen2018hjreachability, hsu2024safetyfilter}.

The deployment implication is that cloud-hosted autonomy should be graded by function. Cloud assistance is well suited to shared perception updates, digital twins, fleet coordination, and non-real-time policy improvement. Safety-critical actuation should remain protected by local monitors, fallback controllers, and verified constraints. This function-level distinction is essential for positioning CCSs within trustworthy robotics and autonomous systems.

\subsection{Transportation and Mobile Autonomy}
In transportation systems, CCSs can aggregate traffic sensors, connected-vehicle data, road-infrastructure signals, and fleet information to support predictive routing, adaptive signal control, and energy-efficient vehicle coordination \citep{smarttransfor2, smarttransfor1, smarttransfor3}. Connected and autonomous vehicles can benefit from cloud-provided situational awareness, such as congestion, weather, road closures, and incident information \citep{smarttransfor4, smarttransfor5}. Recent cloud-enabled MPC and cloud-based data-enabled predictive-control studies show how this idea can be instantiated for vehicle trajectory tracking and mixed traffic-flow regulation \citep{gao2026cloudenabled, chen2026clouddistributed}. The main control challenge is latency and safety: local vehicle control must remain independent for fast dynamics, while cloud services should support longer-horizon coordination and supervisory optimization.

\subsection{Buildings and Human-Centered Automation}
Smart homes and buildings demonstrate CCS value in comfort optimization, energy management, occupancy-aware HVAC control, security monitoring, and remote maintenance \citep{smarthomes1, smarthomes2}. These systems are usually slower than transportation or robotics, making them suitable for cloud-assisted optimization. Their main limitation is privacy: occupancy patterns, behavior, and security-camera data can be sensitive, so privacy-preserving control and access governance are especially important.

\subsection{Healthcare and Assistive Autonomy}
Healthcare monitoring uses wearable and medical IoT devices to collect physiological signals and transmit them to cloud systems for analytics, anomaly detection, and decision support \citep{health1, health2, health3}. From a CCS perspective, the key opportunity is closed-loop or semi-closed-loop care, where monitoring, prediction, alerts, and intervention recommendations are coordinated. The bottlenecks are reliability, medical-data privacy, regulatory compliance, and the need to distinguish clinical decision support from autonomous control.

\subsection{Urban-Scale Autonomous Infrastructure}
Smart cities integrate transportation, air quality, energy, public safety, and infrastructure monitoring \citep{smartcity1, smartcity2}. CCSs can support city-scale optimization by linking local sensing with cloud-based prediction and resource allocation. The challenge is governance as much as control: different agencies, operators, and citizens may share infrastructure, so accountability, auditability, and data minimization are needed alongside technical scalability.

\subsection{Smart Grids and Energy Autonomy}
Smart grids are among the strongest CCS application domains because they require large-scale coordination of distributed energy resources, demand forecasting, fault detection, and renewable integration \citep{smartgrids1, smartgrids2, smartgrids3, smartgrids4}. Cloud resources can support long-horizon optimization and fleet-level analytics, while edge devices maintain fast protection and local stability. The most important research direction is certified hybrid operation: cloud intelligence should improve efficiency without weakening grid resilience during communication loss or cyberattack.

\section{Open Problems and Research Agenda}
\label{sec:research_agenda}
The next stage of CCS research should move from demonstrating cloud-enabled CPS algorithms toward building dependable cyber-physical computing services. The field now has enough conceptual and algorithmic evidence to identify its grand challenges: real-time guarantees under cloud uncertainty, privacy-preserving computation without excessive delay, hybrid cloud-edge-fog allocation, trustworthy AI integration, reproducible benchmarks, and deployment evidence that can satisfy industrial safety expectations.
\subsection{From Cloud Computation to Deployable Control Services}
\begin{figure}[t!]
\centering
\includegraphics[width=4.4in]{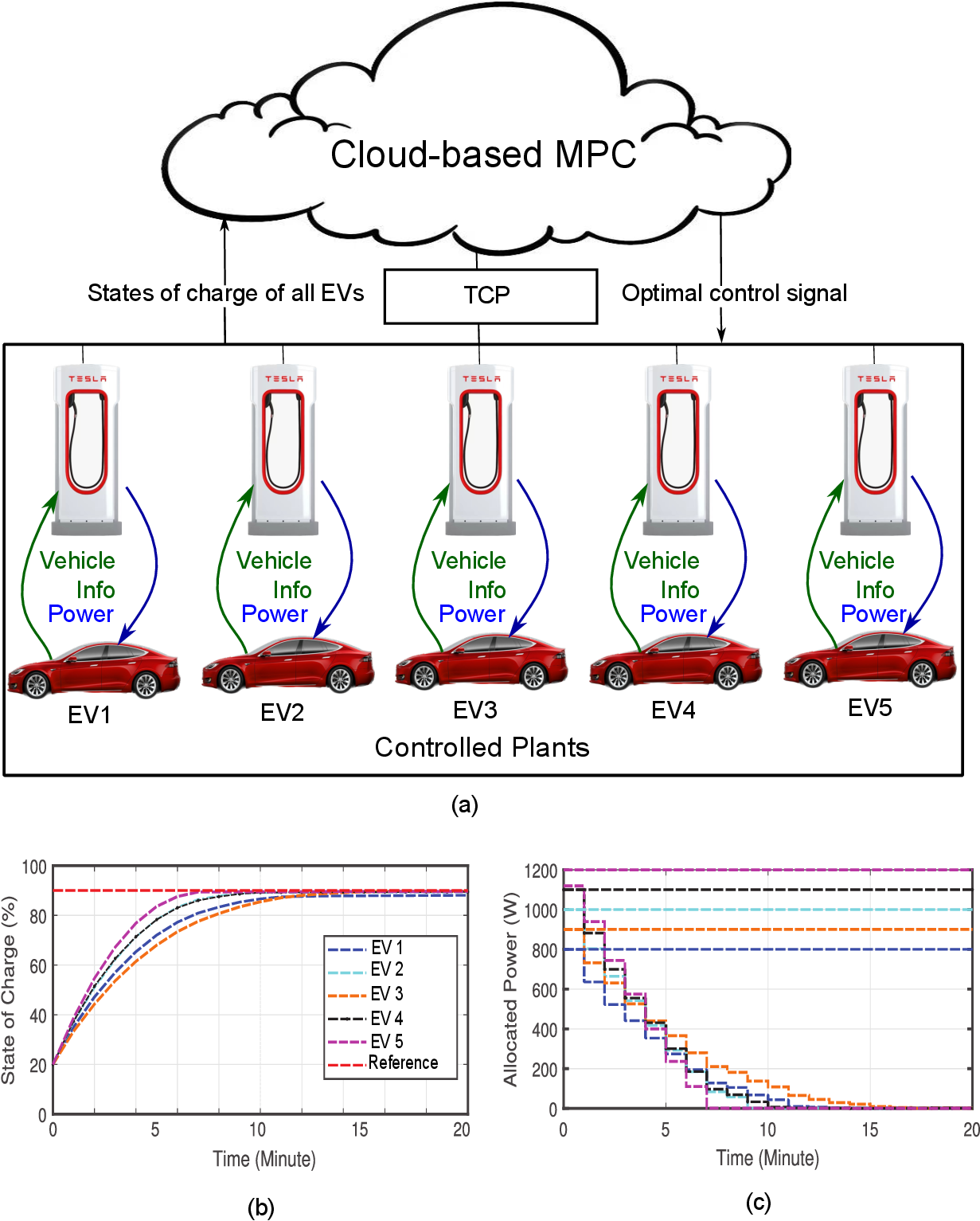}
\caption{Cloud-based control architecture for energy management in EV charging. The figure illustrates a suitable CCS use case: cloud resources execute computational MPC for a relatively slow, resource-coordination problem, while deployment claims still require timing traces, fallback behavior, and outage handling.}
\label{Fig3}
\end{figure}

Future CCSs should not depend on a single remote computing node or assume ideal communication. Cloud-based control techniques must be adapted to practical network settings that include latency, packet loss, server jitter, changing workloads, and cyber-physical attacks. Control activities should be decomposed into subtasks whose placement is justified by timing and safety requirements: local devices handle fast protection loops, fog or edge resources coordinate regional decisions, and the cloud handles long-horizon optimization, learning, and fleet-level analytics. The cloud-based energy-management architecture of \citet{dai2023cloud}, shown in Fig.~\ref{Fig3}, illustrates how parallel multi-block ADMM and cloud containers can support large optimization tasks. The next step is to report end-to-end timing, failure recovery, and fallback behavior with the same rigor as optimization accuracy.
\subsection{Trustworthy AI and Learning-Enabled CCSs}
\begin{figure}[!htbp]
\centering
\includegraphics[width=\textwidth]{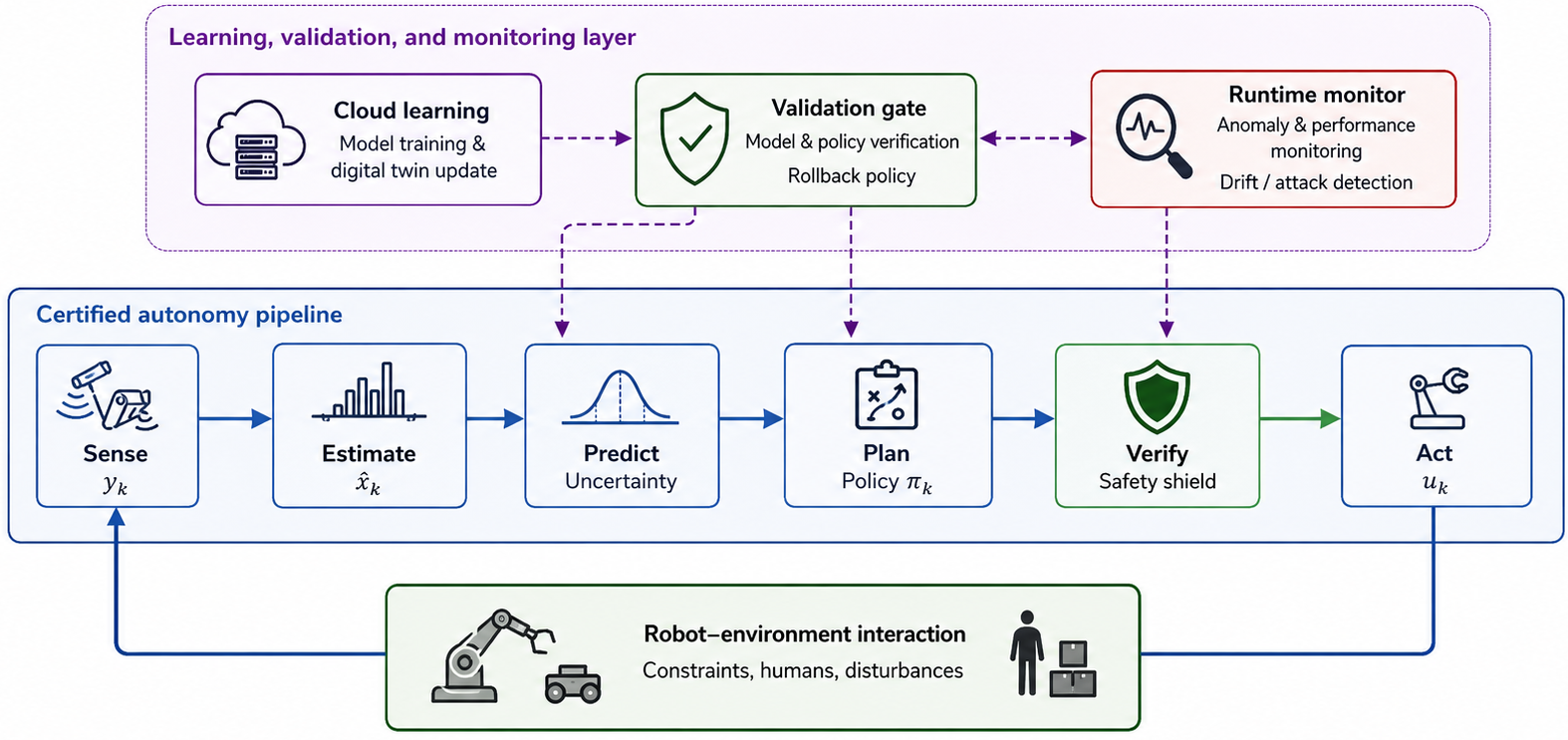}
\caption{Trustworthy AI pipeline for learning-enabled CCSs. Cloud resources can train models and update digital twins, but the feedback loop requires uncertainty estimation, validation, safety shielding, runtime monitoring, and rollback before learned policies are allowed to affect actuation.}
\label{fig:trustworthy_ai_pipeline}
\end{figure}

Artificial Intelligence (AI) and Machine Learning (ML) can expand CCS capabilities through prediction, anomaly detection, adaptive modeling, fault diagnosis, and policy improvement. Cloud infrastructure can train models using distributed GPUs or specialized accelerators and deploy them through scalable services \citep{tuli2023ai, jin2025cloud}. However, learning-enabled CCSs require stronger guarantees than ordinary cloud AI services. A model update can alter closed-loop behavior, so uncertainty quantification, explainability, safe exploration, distribution-shift detection, and rollback mechanisms must become part of the control architecture \citep{brunke2022safelearning, hsu2024safetyfilter}. Federated learning, differential privacy, homomorphic encryption, and secure multi-party computation can reduce data exposure \citep{wang2020edge, alexandru2020secure, huong2021detecting}, but they also change model accuracy, communication load, and real-time feasibility. Future research should therefore treat AI in CCSs as a safety-critical control component rather than as an external analytics layer.
Fig.~\ref{fig:trustworthy_ai_pipeline} summarizes this required trustworthy AI
pipeline. Cloud resources may train models, update digital twins, and improve
prediction or diagnosis, but learned components should influence actuation only
after validation, uncertainty assessment, safety shielding, runtime monitoring,
and rollback planning. This pipeline makes cloud learning compatible with the
closed-loop service view developed throughout this review.

\subsection{Real-Time Privacy-Preserving Control}
Data privacy remains a central concern because cloud controllers may access plant states, operating constraints, user behavior, and proprietary models \citep{xia2024cloud}. Existing privacy-preserving policies often require substantial computation and communication for encryption or decryption, making them difficult to apply to fast feedback loops \citep{paper2, paper5}. Future work should target privacy-preserving algorithms whose sampling-time feasibility is explicitly reported. Important directions include approximate encrypted MPC with certified error bounds, event-triggered encrypted control with leakage analysis, privacy-preserving observers for nonlinear systems, and standardized benchmarks that report privacy strength, control degradation, latency, memory, and energy cost together.

\subsection{Hybrid Cloud-Edge-Fog Architectures and Fallback Certification}
Hybrid architectures are likely to become the dominant practical form of CCSs because cloud-only control cannot meet all latency and resilience requirements \citep{zhan2019future, boiko2024edge, jin2025cloud}. The open research problem is principled task allocation: which computations should remain local, which should move to fog nodes, and which should be centralized in the cloud, guided by the cross-layer trade-offs summarized in Table~\ref{tab:tradeoffs}. Future architectures should include certified fallback controllers, cloud outage protocols, controller handoff conditions, and recovery logic. Stability under switching between local, edge, fog, and cloud controllers must be treated as a first-class design objective rather than an implementation detail.

\subsection{Benchmarks, Standards, and Evidence Maturity}
CCS research still lacks common benchmarks and reporting standards. Studies use different plants, sampling times, threat models, cloud platforms, and performance metrics, making it difficult to compare approaches. A stronger evidence culture would report plant or robot dynamics, sampling period, network-delay traces, packet-loss model, cloud computation-time traces, edge/fog resource constraints, attack model, privacy threat model, encryption and security overhead, control performance, stability or safety metrics, fallback policy, recovery behavior, and deployment cost or resource use. Regulatory and privacy requirements such as GDPR and CCPA also shape what data can be moved to the cloud and how control services should be audited \citep{qi2020efficient, froomkin2022safety}. Benchmark suites and open testbeds would allow the field to move from isolated demonstrations to cumulative engineering knowledge.

A useful CCS benchmark should therefore be multi-layered rather than algorithm-only. It should specify the physical system, the control task, network and cloud traces, the security and privacy threat models, the fallback policy, and the deployment metric. It should evaluate closed-loop behavior under timing uncertainty, cloud outages, cyberattacks, privacy mechanisms, and fallback transitions. Table~\ref{tab:benchmark_requirements} proposes a reporting template that can be used for industrial plants, autonomous vehicles, multi-robot fleets, energy systems, or healthcare monitoring systems. The template is intentionally demanding because the field cannot assess safety-critical cloud control based solely on nominal tracking error or average computation time.

\begin{table}[!ht]
\centering
\caption{Minimum reporting elements for reproducible CCS benchmarks}
\label{tab:benchmark_requirements}
\resizebox{0.97\textwidth}{!}{
\begin{tabular}{p{3.8cm}p{7.0cm}p{7.0cm}}
\toprule
\textbf{Benchmark element} & \textbf{What should be reported} & \textbf{Why it matters for CCS evidence} \\
\midrule
\textbf{Plant or robot model} & Dynamics, constraints, disturbance assumptions, sampling period, sensors, actuators, and safety limits & Enables stability, feasibility, and safety claims to be interpreted \\
\midrule
\textbf{Network and cloud timing} & Uplink/downlink delay traces, packet-loss model, jitter, queueing, cloud solve-time traces, and edge/fog resource limits & Determines whether cloud computation can meet the decision deadline \\
\midrule
\textbf{Security and privacy model} & Attack model, attacker location, encryption or privacy overhead, leakage assumptions, key-management assumptions & Separates secure communication from secure closed-loop behavior \\
\midrule
\textbf{Control performance} & Tracking/regulation metrics, constraint violations, energy or cost, robustness margins, and stability/safety indicators & Connects algorithmic performance to physical consequences \\
\midrule
\textbf{Fallback and recovery} & Local fallback policy, trigger conditions, degraded-mode performance, switchback logic, recovery time, and recovery quality metric & Shows whether the plant remains safe during cloud outage, attack, or stale updates \\
\midrule
\textbf{Deployment and lifecycle} & Testbed or field-trial description, controller-versioning process, audit logs, operator workflow, cost or resource metric, and maintenance plan & Supports claims about operational maturity rather than only simulation feasibility \\
\bottomrule
\end{tabular}}
\end{table}
Table~\ref{tab:grand_challenges} summarizes the grand challenges that structure the next stage of CCS research. These challenges connect real-time cloud control, privacy-preserving feedback, trustworthy AI integration, hybrid architecture design, and evidence standardization into a single deployment-oriented research agenda. The table is intended to make future issues actionable by linking each challenge to its motivation, current bottleneck, promising directions, and expected impact.
\begin{table}[!htbp]
\centering
\caption{Grand challenges for next-generation CCS research}
\label{tab:grand_challenges}
\resizebox{0.97\textwidth}{!}{
\begin{tabular}{p{3.8cm}p{4.4cm}p{4.4cm}p{4.4cm}p{4.4cm}}
\toprule
\textbf{Grand challenge} & \textbf{Why it matters} & \textbf{Current bottleneck} & \textbf{Promising directions} & \textbf{Expected impact} \\
\midrule
\textbf{Real-time cloud control} & Cloud delay and jitter can destabilize feedback loops & Limited end-to-end timing evidence and weak outage models & Predictive buffering, edge fallback, delay-aware MPC, runtime monitoring & Better evidence for industrial and safety-critical CCSs \\
\midrule
\textbf{Privacy-preserving feedback} & Plant data and models may be commercially or personally sensitive & Encryption and privacy tools increase computation and communication cost & Approximate encrypted control, differential privacy with performance bounds, secure aggregation & Cloud services that preserve confidentiality without unacceptable control degradation \\
\midrule
\textbf{Trustworthy AI integration} & AI can improve prediction and diagnosis but may introduce unsafe updates & Limited guarantees under distribution shift and adversarial data & Safe learning, uncertainty quantification, certified rollback, federated learning & Adaptive CCSs with auditable and safe model evolution \\
\midrule
\textbf{Hybrid architecture design} & Cloud, edge, and fog layers have different latency, cost, and resilience properties & Task allocation is often ad hoc & Control-aware orchestration, switching stability analysis, cloud-edge co-design & Better balance among scalability, responsiveness, and resilience \\
\midrule
\textbf{Evidence and standardization} & Industrial adoption requires reproducible performance and risk evidence & Studies use incompatible metrics and assumptions & Shared benchmarks, testbeds, reporting checklists, safety-case templates & Cumulative research progress and clearer certification pathways \\
\bottomrule
\end{tabular}}
\end{table}
The importance of these deployment factors is further contextualized by the landscape of ongoing research. As shown in Table~\ref{tab:ccs_comprehensive}, the evolution of CCSs from early theoretical formulations to secure, robust, and real-time implementations reflects a clear trajectory toward maturity. Each entry highlights a distinct contribution, whether in architecture, domain-specific application, resilience, or secure computation, and also exposes the remaining gap between promising prototypes and broadly deployable CCS infrastructure.
\begin{table}[!htbp]
\centering
\caption{Selected evolution of CCS research from concept toward deployable control services (2012--2026)}
\label{tab:ccs_comprehensive}
\begingroup
\renewcommand{\arraystretch}{1.06}
 \resizebox{0.97\textwidth}{!}{\begin{tabular}{p{2.5cm}p{5cm}p{4cm}p{5cm}p{5cm}}
\toprule
\textbf{Study} & \textbf{Control \& Security Approach} & \textbf{Application Domain} & \textbf{Key Innovations} & \textbf{Challenges \& Future Directions} \\
\midrule
\citet{13-xia2012networked} &
General concept transition from NCS to CCS&
Distributed Cloud-Based MPC &
Predictive control with dynamic node delegation &
Node synchronization, delivery delay compensation \\

\midrule
\citet{givehchi2014control}&
Control-as-a-Service &
Cloud Platforms &
Service-oriented control paradigm &
Standardization needs \\

\midrule
\citet{younis2014access} &
Adaptive access control for on-demand services &
Cloud computing &
Multi-tenant-aware dynamic permission model &
Policy heterogeneity \& insider threat resilience \\

\midrule
\citet{xia2015cloud} &
Theoretical Foundations &
General Theory &
Identified fundamental CCS challenges &
Stability under delays \& jitter \\

\midrule
\citet{LEI2016324} &
Resilient MPC &
Industrial Automation &
Packet loss-tolerant predictive control &
QoS in shared clouds \\

\midrule
\citet{kobara2016cyber} &
Secure Feedback Control &
Industrial IoT &
Lightweight encryption schemes &
Key management complexity \\

\midrule
\citet{liu2017predictive} &
Cloud Predictive Control + delays and data dropouts &
Multi-agent Systems in robot swarms &
Delay-compensated consensus with distributed cloud nodes &
Stability under time-varying network constraints \\

\midrule
\citet{1-ali2018secure} &
DDoS-Resilient Switching Control &
CCS-based control systems and UAVs &
Mitigation and detection under cloud attack &
Real-time stability during network outage\\

\midrule
\citet{zhan2019future} &
Cloud-Fog Hybrid Control &
IoT Systems &
Incentive-based fog architecture for control offloading &
Latency, congestion in hybrid networks \\

\midrule
\citet{yang2019predictive} &
Quantized Predictive Cloud Control under DoS attacks &
Networked Multiagent Systems &
Quantized predictive control with DoS-resilient packet loss handling &
Quantization-induced instability \& persistent attack resilience \\

\midrule
\citet{li2020integrated} &
Dual Channel-Aware Scheduling with PBMPC &
Wireless Cloud Control Systems &
Game-theoretic sensor scheduling + prioritized cloud-side controller selection &
Convergence-stable under dual-channel SAPs and prediction horizon coupling \\

\midrule
\citet{skarin9683307} &
Explicit MPC with Verified Recovery Paths &
Elastic Cloud-Control Systems &
Resilient dual-controller design with error-signal recovery via predictive fallback &
Feasibility loss under short horizons \& cost-aware switchback to nominal MPC \\

\midrule
\citet{Yuan9987498} &
Application-Layer Coding with RaptorQ Fountain Codes &
Industrial Cloud Control Systems &
Low-latency, reliable transmission using adaptive fountain codes for time-critical industry &
Latency-cost tradeoff tuning \& decoding overhead under variable erasure rates \\

\midrule
\citet{2-ali2023chaos} &
Chaos-Encrypted Polar Coding &
Network-Oriented Cloud Control Systems &
3D-Lorenz-based internal encryption for polar-coded 16QAM-OFDM with secure subcarrier mapping &
Key synchronization, scalability for bulk encryption \\

\midrule
\citet{lyu2024modeling} &
Switching Control under Uncertainty &
Industrial IoT &
Quadratic optimal controller for delay-tolerant CCS &
Network-induced model switching \& stability analysis \\

\midrule
\citet{zhorfei10879591} &
Delay-Dependent Cloud MPC &
Vehicular Platoon Systems &
HDD-based string-stable platooning prediction under bidirectional delay &
Consensus loss under high delay \& model mismatch\\
\bottomrule
\end{tabular}}
\endgroup
\end{table}
\section*{Key Findings}
\begin{enumerate}
\item CCSs are best understood as cloud-edge-fog cyber-physical service infrastructure, not as conventional NCSs with stronger remote computers.
\item Delegating intelligence changes the computing problem because sensing, estimation, optimization, learning, privacy, security, audit, and recovery may be distributed across layers, organizations, software stacks, and data-governance boundaries.
\item Cloud assistance is strongest for long-horizon optimization, high-dimensional MPC, digital twins, fleet coordination, model learning, and audit services; fast safety-critical loops usually require local or edge authority.
\item Timing-aware task allocation, middleware QoS, runtime assurance, and verified fallback are deployment conditions rather than optional implementation features.
\item Security and privacy mechanisms must be evaluated by closed-loop service consequences, including delay, estimation accuracy, command integrity, stability margins, recovery behavior, and accountability.
\item Evidence maturity remains the limiting factor for robotics and safety-critical autonomous CPSs. Simulation-only studies are useful but cannot support strong deployment claims without timing traces, attack models, fallback validation, operator workflow, and lifecycle evidence.
\end{enumerate}

\section{Conclusions}
CCSs are emerging as a new cyber-physical computing layer for trustworthy autonomous CPSs. Their importance does not lie only in moving computation from a local processor to a remote server. It lies in changing where sensing, feedback, estimation, optimization, learning, privacy protection, security monitoring, governance, and recovery are allowed to reside. This change creates opportunities for scalable MPC, fleet coordination, digital twins, secure monitoring, and learning-enabled autonomy, but it also exposes physical systems to risks related to cloud timing, service model updates, privacy, and accountability. The decisive bottleneck for CCSs is the trustworthy operation of closed-loop services. Faster cloud solvers and larger data sets are valuable only when coupled with timing-aware task allocation, dependable middleware, local or edge fallback, privacy-preserving computation, integrity protection, availability under attack or outage, runtime assurance, and deployment evidence. A cloud-assisted controller may be promising in simulation; a deployable CCS must show how it behaves when packets are late, a solver is interrupted, a model update is wrong, a service is unavailable, or an authorized command is unsafe. The future of CCSs in robotics, industrial automation, transportation, energy, healthcare, buildings, and smart infrastructure will depend on co-design. Algorithms, cloud-edge-fog architecture, security mechanisms, AI pipelines, middleware, operator workflows, and safety cases must be developed together. If that synthesis is supported by credible evidence, CCSs could become an enabling infrastructure for trustworthy autonomy: not a replacement for local feedback, but a disciplined computing service layer that extends what physical systems can optimize, learn, coordinate, and verify.

\section*{Declaration of competing interest}
The authors declare that they have no competing financial interests or personal relationships that could influence the work.

\section*{Acknowledgments}
This work was supported in part by the Beijing Natural Science
Foundation Haidian Original Innovation Joint Fund Project under Grant L252035, in part by the National Natural Science Foundation of China under Grant U25A20460, in part by the National Natural Science Foundation of China for International Young Scientists under Grant W2533176, in part by the Beijing Natural Science Foundation under Grant IS25064, in part by the Joint Fund Project of Shandong Provincial Natural Science Foundation under Grant ZR2025LZH001, in part by the Henan Provincial Talent Program under Grant 264000510006 and in part by the Henan Postdoctoral Foundation under Grant HN2026058.

\section*{Data availability}
Data will be available upon
request.
\bibliographystyle{elsarticle-harv}
\bibliography{mybib}

\end{document}